\documentclass{article}

\usepackage[breaklinks]{hyperref}
\usepackage{lineno}
\modulolinenumbers[5]

\usepackage[utf8]{inputenc}
\usepackage{amsmath}
\usepackage{psfrag}
\usepackage{subfig}
\usepackage{color}
\usepackage{empheq}
\usepackage{tikz}
\usepackage{verbatim}
\usepackage{textcomp}
\usepackage{siunitx}
\usepackage{float}

\begin{document}

\title{Numerical simulation of the viral entry into a cell driven by receptor diffusion}

\author{\\ T. Wiegold$^{1,}$\thanks{Corresponding author, email address: tillmann.wiegold@tu-dortmund.de} , Sandra Klinge$^{2}$, R. P. Gilbert$^3$, G. A. Holzapfel$^{4,5}$}

\date{\normalsize{$^{1}$ Institute of Mechanics, TU Dortmund University,Leonhard-Euler-Strasse 5, 44227 Dortmund, Germany\\
$^2$ Chair of Structural Mechanics and Analysis, TU Berlin, Strasse des 17. Juni 135, 10623 Berlin, Germany\\
$^3$ Department of Mathematical Sciences, University of Delaware, 312 Ewing Hall, Newark, Delaware 19716, USA\\
$^4$ Institute of Biomechanics, Graz University of Technology, Stremayrgasse 16/2, 8010 Graz, Austria\\
$^5$ Department of Structural Engineering, Norwegian University of Science and Technology (NTNU), 7491 Trondheim, Norway}}

\maketitle
\vspace{-5mm}
\begin{center}The formal publication of this article can be found via:\\
https://doi.org/10.1016/j.camwa.2020.12.012\end{center}
\begin{abstract}
The present study focuses on the receptor driven endocytosis typical of viral entry into a cell.
A locally increased density of receptors at the time of contact between the cell and the virus is necessary in this case. The virus is considered as a substrate with fixed receptors on its surface, whereas the receptors of the host cell are free to move over its membrane, allowing a local change in their concentration. In the contact zone the membrane inflects and forms an envelope around the virus. The created vesicle imports its cargo into the cell. This paper assumes the diffusion equation accompanied by  boundary conditions requiring the conservation of binders to describe the process. Moreover, it introduces a condition defining the energy balance at the front of the adhesion zone. The latter yields the upper limit for the size of virus which can be engulfed by the cell membrane. The described moving boundary problem in terms of the binder density and the velocity of the adhesion front is well posed and numerically solved by using the finite difference method. The illustrative examples have been chosen to show the influence of the process parameters on the initiation and the duration of the process.
\end{abstract}
~

\newpage

\section{Introduction}
\label{Introduction}
The intense study of cell mechanisms has provided an important insight into the uptake of various substances into a cell including viruses. Diagnosis and therapy of diseases has reached a state in which nanomedicine concerned with devices of nanoscale size is applied. With these, they can deliver low molecular mass compounds, proteins and recombinant DNAs to focal areas of disease. Some examples are polymeric micelles, quantum dots, liposomes, polymer-drug conjugates, dendrimers, biodegradable nanoparticles, silica nanoparticles, etc. researched in laboratories, undergoing preclinical development, or  already used in hospitals \cite{davis2010, Tasciotti:2008}. In addition, numerical methods in biomechanics and biomathematics become more relevant. A variety of numerical methods ranging from multiscale finite elemente \cite{klinge2010} and isogeometric shell formulations \cite{duong2017} to relaxation dynamics \cite{arroyo2009} find their application in biomechanics.

The cell is surrounded by a plasma membrane which acts as the interface between the cell and its surrounding environment. However, the membrane is not absolutely impermeable and a transport of particles through the membrane is still possible. Among many different mechanisms, the most common process for this purpose is the so-called endocytosis \cite{Geoffrey:2000}. The main focus of the investigation of endocytosis has so far been on clathrin-mediated endocytosis (CME). During the CME, proteins create clathrin-coated pits which eventually build whole vesicles \cite{Sorkin:2004, Schmid:2007}.

Various aspects of the endocytosis process are investigated in different chemical and biochemical contexts. Amongst others, the total internal reflection illumination with fluorescence correlation spectroscopy is used to measure ligand-receptor kinetic dissociation rate constants \cite{lie1}. Trafficking phenomena are studied based on internalization experiments utilizing multiscreen assay systems \cite{fallon2000}. Furthermore, biomimetic systems of lipid vesicles or supported bilayers with a variety of binder molecules deal with soft adhesion mediated by mobile binders as shown in \cite{nam2007, fenz2017}. The behavior and stability of adhesion complexes are also addressed by using cell doublet \cite{tozeren1989,engl2014} and  vesicles adhered to supported bilayers \cite{smith2008}.

The experimental progress has parallely led to multiple theoretical models. A mathematical framework, based on probabilities for binding rates, introducing a random and a sequential driving mechanism for the receptors is provided in \cite{gib1}. Discrete stochastic models for specific receptor-ligand adhesion as well as non-equilibrium continuum models for the competition between different modes of junction remodeling under force are developed in \cite{kaurin2018}, whereas a statistical thermodynamic model of viral budding is presented in \cite{tzlil2004}. Several analytical models for the endocytosis process utilize the description of the Stefan problem, \cite{fre1,gao2005} and propose a solution relying on the error and complementary error functions. These methods are particularly applied to the HIV-Virus \cite{sun2006} and Semliki Forest virus \cite{gao2005} but also for the uptake of nanoparticles \cite{yi2017}. Alternatively, Tseng and Huang \cite{tseng2014} use the immersed  boundary method to simulate the endocytosis  and to investigate the resistance of the water film in the contact area.

The current contribution uses the model presented in \cite{fre1,gao2005} as a basis, however, it proposes an alternative form of the Stefan supplementary condition where the focus is only set on the energetic aspects of the front itself, and a consideration of the dissipation  associated  with the receptor transport along the  cell membrane  is not needed. Different from the previous models, the new formulation yields the upper limit for the size of virus  able to enter the cell. Another focus of the paper is the purely numerical  solution of the moving boundary problem where no additional  assumptions typical of analytical solutions such as  the speed factor  \cite{fre1,gao2005} are necessary. The chosen interpretation of the problem is advantageous for the fast simulation of various scenarios regarding the process parameters  and their influence on the initiation and duration  of the process. As a final objective, the paper also introduces the notion of cooperativity \cite{Krobath2009,Weikl:2009} into the numerical model and performs a more sophisticated  study of the effective receptor distribution on a virus and of its influence on the viral entry into a cell.

This contribution is structured as follows: A general overview of the uptake process is summarized in Sect. \ref{Description of the uptake process}, whereas Sect. \ref{Process characterization} recapitulates the main aspects of the free energy characteristic for the endocytosis process. After this introductory part, Sect. \ref{Boundary conditions} focuses on the definition of boundary and supplementary conditions accompanying the driving diffusion differential equation. The boundary conditions define the flux balance, and supplementary condition represents the energy balance at the adhesion front. Subsequently, Sect. \ref{Numerical implementation and results for 1D case and helical viruses} discusses the numerical implementation  for the 1D case and helical viruses, which is followed by a non-dimensional description of the problem in Sect. \ref{Nondimensionalization}. The paper also deals with  a common case of a spherical virus (Sect.~\ref{Rotationally symmetric case - Spherical virus}) and extends the basic model by introducing the notion of cooperativity (Sect. \ref{Cooperativity}). Finally, Sect. \ref{Discussion} gives a comprehensive overview of results and provides a comparison with the experimental observations and numerical results from the literature. The paper finishes with a conclusion and an outlook.

\section{Description of the uptake process}
\label{Description of the uptake process}
In a basic view of mechanical adhesive contact between elastic surfaces, two phenomena which have a considerable influence on the underlying process counteract each other.
A reduction in the free energy when surfaces with bonding potential come into contact benefits the process, whereas an increase in free energy due to elastic deformation required to fit their shapes counteract the process.
In the classic Hertzian theory of elastic deformation \cite{Johnson:1985}, two bodies coming into contact deform in the contact area in such a way that they perfectly fit.
According to this approach any surface interactions such as Van der Waals forces, which are induced by charge polarization in electrically neutral molecules in close proximity, are excluded.
However, these 'non-material' effects have a significant influence on the direct contact interaction \cite{bell1978}.
This is illustrated by considering small elastic objects consisting of crystalline materials processed in a controlled environment.
Such crystals show the appearance of unfulfilled or dangling chemical bonds distributed over a free surface.
Bringing such objects into contact reduces the free energy of the system by forming bonds between the two surfaces.
The objects joined in this way will not separate without additional work.
Hence, not only compressive traction due to bulk elasticity but also an adhesive or tensile traction contributes to the contact \cite{fre1}.
General investigations of the mechanisms on adhesive contact are presented in \cite{Johnson:1971, derjaguin1975, maugis1992, kim1998}.

The same effects, both attractive and resisting interactions, appear in the adhesive contact of biological cells.
However, due to their characteristic properties compared to engineering materials, significant differences occur in this case.
Having a remarkably lower elastic modulus than engineering materials weakens the influence of the effect of elastic energy variations during contact \cite{fre1}.
Furthermore, cells are characterized by a fluid-like in-plane behavior\cite{sauer2017}.
This enables the receptors of the cell to move within its membrane, enabling new methods of incorporating free energy variations in the modeling of adhesive contact.

For a long time, electron microscopy has been used  to provide valuable insight into the architecture of viruses \cite{dales1962}. Furthermore, fluorescent-labeling of viruses and cellular structures combined with  fluorescence microscopy yield more dynamic information for the tracking of a single virus in live cells \cite{brandenburg2007}. In order to obtain 3D geometrical and distributional information, electron tomography has also shown to be a powerful tool \cite{barrow2013}. Thus, a large amount of information on the architecture of viruses is already available.
For example,  a spike protein density of approximately $ 2800 \,\,\si{\micro}$m$^{-2}$ is identified for  Sars-Cov  \cite{beniac2006} and   the value of approximately $ 5200\,\, \si{\micro}$m$^{-2}$ for an Alphavirus  \cite{tzlil2004}. With regard to the geometry, the investigations have shown that the helical and spherical virus forms are predominant \cite{crick1956}.

\begin{figure}
	\centering
	\includegraphics[width = 1.\textwidth]{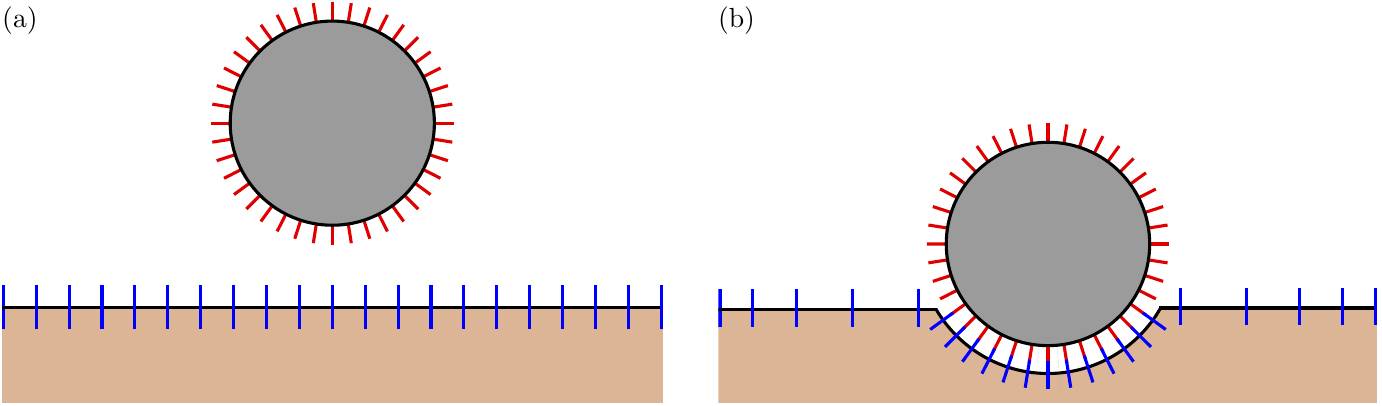}
	\caption{(a) Initial configuration of the cell surface and the virus in a 2D setup. (b) State during the uptake where the virus is partially connected to the cell.}
	\label{fig:P1_1}
\end{figure}

In order to depict the process of viral entry into a cell, the situation presented in Fig. \ref{fig:P1_1} is considered. This 1D situation is suitable to simulate the endocytosis of helical virus into a cell. However, an extension to the rotationally symmetric case and to the simulation of a spherical virus is straightforward (Sect.~\ref{Rotationally symmetric case - Spherical virus}). In the initial state, the virus has not yet reached the cell surface (Fig. \ref{fig:P1_1}a). Upon first contact, the virus gradually connects to the cell (Fig. \ref{fig:P1_1}b). In order to establish a connection between the virus and the cell, a generic repulsion between their surfaces needs to be overcome.

The connection by binding receptors of the virus to receptors of the cell reduces the internal energy of the system. Upon completing a single receptor-ligand bond, the internal energy is reduced by $k\,T\,C_{\text{b}}$, where $k$ is the Boltzmann constant, $T$ the absolute temperature and $C_{\text{b}}$ the binding energy coefficient. 

The quantity driving the uptake process is receptor density $\xi$.
Initially, the density of receptors on the cell surface amount to $\xi_0$ and the corresponding counterpart, the receptor density on the virus surface, amounts to $\xi_{\text{rv}}$. In general, it holds that the density of receptors on the virus is larger than the one on the cell surface and that the virus receptors are fixed, whereas the receptors of the cell are free to move across the membrane. Upon contact, receptors of the cell diffuse over the surface, connect to the receptors of the virus and build an envelope around the virus. At the end of the process the envelope is closed over the virus which has fully entered the cell.

As opposed to metallic or covalent bonds, the bonds created during biological adhesion are relatively weak.
Since the cell receptor density is the lower one, in general, it dictates the amount of reduction in the internal energy. Typically, the resisting potential due to generic repulsion exceeds the reduction in internal energy of the initial configuration of the system for a unit area of the membrane at $\xi_0$.
Therefore, additional influences facilitate the creation or dissolving of chemical bonds.
Possible influences are catalytic agents, small temperature changes and small mechanical forces.
It appears that a local change in receptor density is necessary in order to create an adhesion zone between the virus and the cell.
An increasing local receptor density results in a greater reduction in the free energy by completion of each additional bond.
When the cell receptors and the virus receptors are close to each other, a permanent interaction is present due to thermal stimulation.

In the framework of chemical rate theory two distinct cases are differentiated \cite{Dembo:1988}.
In the area where $\xi < \xi_{\text{eq}}$ holds, the rate of bond breaking exceeds the rate of bond forming, so that no adhesive contact is established.
In the area where $\xi > \xi_{\text{eq}}$ holds, the rate of bond forming exceeds the rate of bond breaking, and a strong adhesive contact is established.
Condition $\xi = \xi_{\text{eq}}$ defines  the chemical equilibrium of the bonding reaction as well as the lower limit for the adhesion to start.
Whereas it is known that $\xi_{\text{eq}} > \xi_{0}$ in the case of cell-virus contact, different assumptions can be made for the exact value of $\xi_{\text{eq}}$.
In a limiting case, the chemical equilibrium requires all receptors of the virus to be bonded to the cell membrane such that $\xi_{\text{eq}} = \xi_{\text{rv}}$.
It is more likely to expect that the equilibrium density $\xi_{\text{eq}}$ is in the range $[\xi_{0},\xi_{\text{rv}}]$ and that it might vary during the process.
Some of these aspects are considered in Sect. \ref{Cooperativity} on cooperativity, an effect significantly influencing the exact value of $\xi_{\text{eq}}$.
However, the initial study is performed for the most restrictive case $\xi_{\text{eq}} = \xi_{\text{rv}}$, which does not influence the generality of the model.

\begin{figure}
	\centering
	\includegraphics[width = 1.\textwidth]{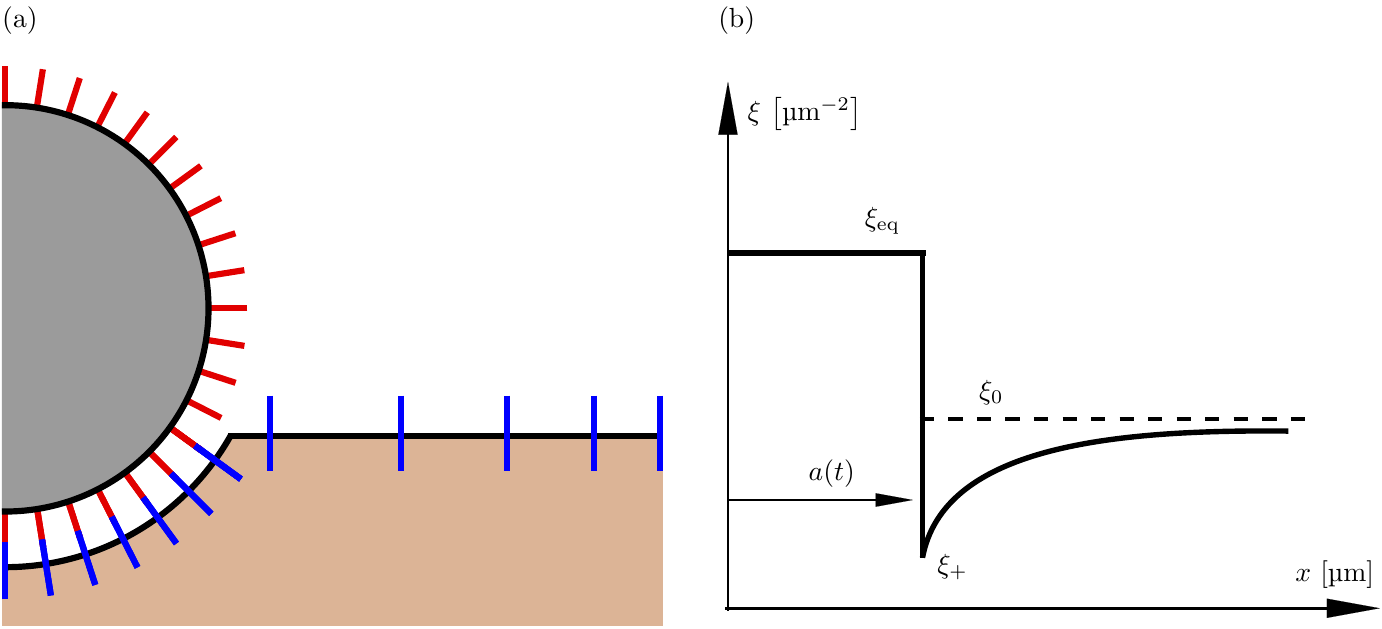}
	\caption{(a) Schematic diagram for the receptor distribution over the cell and virus surface. (b) Typical density profile.}
	\label{fig:P1_2}
\end{figure}

As an illustration, a schematic distribution of receptors over the cell and the virus is depicted in Fig. \ref{fig:P1_2}a, whereas the corresponding density profile is shown in Fig. \ref{fig:P1_2}b.
In the adhesive zone, the receptor density is constant and amounts to $\xi_{\text{eq}}$.
The density starts to grow outside the adhesive zone and, only when far away from it, moves towards the initial density value of the cell $\xi_0$.
Since the size of the cell is magnitudes larger than the virus, we assume the receptor density far away from the adhesion front to stay constant $\lim\limits_{x \rightarrow \infty}{\xi(x,t)} = \xi_0$.
Consequently, the flux $j$ of the receptors vanishes so that $\lim\limits_{x \rightarrow \infty}{j(x,t)} = 0$.

One more peculiarity of diagram \ref{fig:P1_2}b is the front of the adhesion zone where a jump of the receptor density occurs.
The position of the front is denoted by a time dependent function $a(t)$.
The discontinuous profile is expected since the equilibrium density is high and certainly larger than $\xi_{0}$, whereas the receptor density ahead of the front must be lower than $\xi_{0}$ in order to stipulate the receptor diffusion.
The values typical of the adhesion front play an essential role in the model to be described, and are denoted by the subscript $+$ in the subsequent text.
For example, $\xi_{+}$ denotes the receptor density at the front.

The previous explanation shows that the whole process is regulated by the diffusion of receptors over the cell surface and their gathering in the adhesion zone.
Accordingly, the motion of the receptors will be described by the diffusion differential equation
\begin{linenomath*}
	\begin{equation}
		\label{conservation}
		\frac{\partial \xi}{\partial t} = -\dfrac{\partial j}{\partial x}
	\end{equation}
\end{linenomath*}
which states that the change of receptor density in time has to be equal to the negative spatial change in the flux.
Furthermore, following Fick's first law, the receptor flux $j$ is proportional to the gradient of density, i.e.
\begin{linenomath*}
	\begin{equation}
		\label{flux1}
		j = -m\,\frac{\partial \xi}{\partial x},
	\end{equation}
\end{linenomath*}
such that the implementation of Eq.~\eqref{flux1} into Eq.~\eqref{conservation} leads to the alternative expression of the diffusion equation
\begin{linenomath*}
	\begin{equation}
		\label{diff}
		\frac{\partial \xi}{\partial t} = m\frac{\partial^2 \xi}{\partial x^2}.
	\end{equation}
\end{linenomath*}
This equation defines the relation between the temporal and the spatial changes of the receptor density weighted by the mobility parameter $m$.
Its evaluation gives insight into the evolution of receptor density for every point in front of the adhesion zone $a(t) < x < \infty$.
Equation (\ref{diff}) is a partial differential equation of second order and requires additional boundary conditions in order to determine the complete particular solution.
These two conditions will be defined in the upcoming sections.

\section{Process characterization}
\label{Process characterization}
In order to define the free energy characteristic of the simulated process, the system including a large number of receptors is treated analogously to the case of an ideal gas with a large number of non-interacting particles $N$.
In such a case, the entropy of a single particle, belonging to a system in equilibrium, is expressed by $k\,\ln[(A/\Lambda^2)\,(c/N)]$ \cite{fre1}.
Here, $A$ is the considered surface, $N/A$ is the areal density $\xi$, $c$ is a numerical factor and $\Lambda$ a molecule length scale \cite{dill2012}.
However, the latter two quantities ($c$ and $\Lambda$) do not play any role for the description of our process since it does not depend on the absolute entropy but on its change.
This change is described with respect to the initial state of the cell with uniform density $\xi_{0}$ which is chosen to be the reference state.
The relative entropy of a single receptor at density $\xi$ is then described by its difference to the reference state and is calculated according to \cite{fre1} by
\begin{linenomath*}
	\begin{equation}
		\label{entropy}
		k\,\ln\left(\frac{c}{\xi\Lambda^2}\right)-k\,\ln\left(\frac{c}{\xi_0\Lambda^2}\right) = - k\,\ln\left(\frac{\xi}{\xi_0}\right).
	\end{equation}
\end{linenomath*}
With the expression for one receptor at hand, and assuming that the gradient in the distribution is small and that the local distribution is indistinguishable from an equilibrium distribution at local density, the free energy $E_\text{e}$ per unit area of membrane surface associated to the receptor distribution at absolute temperature $T$ turns into
\begin{linenomath*}
	\begin{equation}
		E_\text{e} = k\,T\,\xi\, \ln\left( \frac{\xi}{\xi_0}\right).
	\end{equation}
\end{linenomath*}
Moreover, the chemical potential $\chi$ is defined as the local change in the free energy per receptor,
\begin{linenomath*}
	\begin{equation}
		\label{chi}
		\chi(x,t) = \frac{\partial E_\text{e}}{\partial \xi} = k\,T\,\left[ \text{ln}\left( \frac{\xi}{\xi_0} \right) +1\right].
	\end{equation}
\end{linenomath*}
Finally, the mean receptor speed is assumed to be proportional to the spatial gradient of the chemical potential, i.e.
\begin{linenomath*}
	\begin{equation}
		\label{speed}
		v_\text{r} = -\frac{m}{k\,T}\frac{\partial \chi}{\partial x} = -\frac{m}{\xi}\frac{\partial \xi}{\partial x},
	\end{equation}
\end{linenomath*}
where the motion of the receptors is controlled by the mobility parameter $m$.

\section{Boundary and  supplementary conditions}
\label{Boundary conditions}
The full description of the adhesion front motion relies on a problem formulation including the differential equation \eqref{diff} along with two boundary conditions describing the flux at the ends of the  domain  and along with a supplementary condition.

\subsection{Flux balance}
The boundary conditions on the unbound area are concerned with the quantitative description of the flux of receptors through the adhesive front.
Following the Leibniz integration rule of the global form, this condition is derived from Eq.~\eqref{conservation} as
\begin{linenomath*}
	\begin{equation}
		\label{ic1a}
		(\xi_{\text{eq}} - \xi_+)\,v_+ + j_+ = 0,
	\end{equation}
\end{linenomath*}
or by using Fick's first law as
\begin{linenomath*}
	\begin{equation}
		\label{ic1b}
		(\xi_{\text{eq}}-\xi_+)\,v_+ - m\left[ \frac{\partial \xi}{\partial x} \right]_+ = 0.
	\end{equation}
\end{linenomath*}
Here, the first term denotes the amount of receptors required for the advancement of the front, and the second term denotes the amount of receptors provided by the flux.
Both previous formulations depend on the front velocity defined in terms of the front position $v_+ = \frac{\text{d}\,a}{\text{d}\,t}$.
Equation \eqref{ic1b} is consistent with the assumption \eqref{speed}, which can easily be shown  as follows.
First, the flux is assumed to be proportional to the receptor distribution $\xi$ and the mean receptor velocity $v_\text{r}$:
\begin{linenomath*}
	\begin{equation}
		\label{flux}
		j = \xi\, v_\text{r}.
	\end{equation}
\end{linenomath*}
By incorporating Eq. \eqref{speed} into Eq. \eqref{flux}, the flux turns into
\begin{linenomath*}
	\begin{equation}
		\label{flux2}
		j = -\frac{m}{k\,T}\,\xi\,\frac{\partial \chi}{\partial x} = -m\frac{\partial \xi}{\partial x},
	\end{equation}
\end{linenomath*}
as predicted by Fick's first law.

\subsection{Energy balance}
The supplementary condition  is provided by considering the energetic aspects of the front motion.
The change of the receptor distribution as well as of the membrane shape leads to several contributions to the free energy of the system.
However, the crucial observation is that the difference in the energy ahead of and behind the front results in the front motion, which is expressed as follows, 
\begin{linenomath*}
	\begin{equation}
		\label{energy1}
		E^- - E^+ = E_{\text{kin}}.
	\end{equation}
\end{linenomath*}
Here, $E^-$ denotes the energy behind the adhesion front, $E^+$ is the energy ahead of the front and $E_{\text{kin}}$ is the kinetic energy of the front itself.

The term related to the energy behind the front is built of three contributions, all denoted by superscript $-$,
\begin{linenomath*}
	\begin{equation}
		\label{behind}
		E^- = E_{\text{b}}^- + E_{\text{e}}^- + E_{\kappa}^-.
	\end{equation}
\end{linenomath*}
These terms have the following physical meaning: $E_{\text{b}}^-$ is the energy related to the binding of receptors, $E_{\text{e}}^-$ the energy related to the entropy and $E_{\kappa}^-$ the energy related to the bending of the membrane.
The reduction in the free energy due to the binding of receptors of the cell to receptors of the virus is defined as follows
\begin{linenomath*}
	\begin{equation}
		E_{\text{b}}^- = -k\,T\,C_{\text{b}}\,\xi_{\text{eq}}.
	\end{equation}
\end{linenomath*}
This term is proportional to the reduction of energy caused by a single bond between two receptors $-k\,T\,C_{\text{b}}$, and to the total amount of created bonds $\xi_{\text{eq}}$ dictated by the virus.
As stated in \cite{fre1}, the binding energy coefficient $C_{\text{b}}$ typically takes values in the range $5 < C_{\text{b}} < 35$.
The second term describes the energy associated with the entropy of receptors
\begin{linenomath*}
	\begin{equation}
		\label{entropy_behind}
		E_{\text{e}}^- = k\,T\,\xi_{\text{eq}}\,\text{ln}\left(\frac{\xi_{\text{eq}}}{\xi_0}\right),
	\end{equation}
\end{linenomath*}
which is required to bring the density from its reference value $\xi_0$ to the density of the virus $\xi_{\text{eq}}$. This term will result in an increase in the free energy since it holds $\xi_0 < \xi_{\text{eq}}$. The third term of (\ref{behind}) is concerned with the bending of the membrane caused by the geometry of the virus
\begin{linenomath*}
	\begin{equation}
		E_{\kappa}^- = \frac{1}{2}\,k\,TB\,\kappa^2.
	\end{equation}
\end{linenomath*}
Here, a simplified case is considered corresponding to the theory of the bending of a plate.
Factor $B$ represents the non-dimensional numerical parameter for the bending stiffness, which is in the range of 10 to 30 and $\kappa = 1/R$ represents the curvature, which is constant for a spherical virus and which depends on the radius of the virus $R$.
Thus, the whole energy behind the front is then defined by the expression
\begin{linenomath*}
	\begin{equation}
		\label{E-}
		E^- = -k\,T\,C_{\text{b}}\,\xi_{\text{eq}} + k\,T\,\xi_{\text{eq}}\,\text{ln}\left(\frac{\xi_{\text{eq}}}{\xi_0}\right) + \frac{1}{2}\,k\,TB\,\kappa^2.
	\end{equation}
\end{linenomath*}
In the second step we consider the energy ahead of the front, denoted by superscript $+$.
Binding between the cell and the virus exclusively takes place in the area behind the front and thus does not have any influence on the energy ahead of the front.
However, corresponding parts $E_{\text{e}}^+$, the energy related to the entropy and $E_{\kappa}^-$, the energy related to the curvature of the membrane remain available.
Moreover, a term $E_\text{v}^+$, the energy related to the motion of receptors also has to be taken into consideration.
In summary, the following terms can be counted ahead of the front:
\begin{linenomath*}
	\begin{equation}
		\label{ahead}
		E^+ = E_{\text{e}}^+ + E_{\kappa}^+ + E_{\text{v}}^+.
	\end{equation}
\end{linenomath*}
In the present contribution, we assume that the curvature behind the front is much smaller than the one caused by the contact with the virus.
This justifies the assumption of a vanishing influence to the energy associated to the bending of the membrane
\begin{linenomath*}
	\begin{equation}
		\label{Ekappa+}
		E_{\kappa}^+ = 0.
	\end{equation}
\end{linenomath*}
The energy ahead of the front related to the entropy is expressed in the same way as the energy behind the front as
\begin{linenomath*}
	\begin{equation}
		\label{Ee+}
		E_{\text{e}}^+ = k\,T\,\xi_+\text{ln}\left( \frac{\xi_+}{\xi_0}\right).
	\end{equation}
\end{linenomath*}
It describes the energy needed in order to bring the initial receptor density $\xi_0$ to the value $\xi_{+}$. Contrary to the contribution behind the front, this term results in a reduction of the free energy since $\xi_0 > \xi_{+}$.
The contribution due to the motion of the receptors reads
\begin{linenomath*}
	\begin{equation}
		\label{Ev+}
		E_{\text{v}}^+ = \frac{1}{2}\,m_{\text{r}}\,\xi_+ \,v_\text{r}^2 = \frac{1}{2}\,m_{\text{r}}\,\frac{m^2}{\xi_+} \,\left(\frac{\partial \xi_{+}}{\partial x}\right)^2.
	\end{equation}
\end{linenomath*}
It represents the kinetic energy of all receptors ahead of the front moving towards the front with their corresponding velocity $v_\text{r}$ and the cell receptor mass $m_{\text{r}}$.
With Eqs. \eqref{Ekappa+} - \eqref{Ev+}, the total energy ahead of the front is defined by
\begin{linenomath*}
	\begin{equation}
		\label{E+}
		E^+ = k\,T\,\xi_+\text{ln}\left( \frac{\xi_+}{\xi_0}\right) + \frac{1}{2}\,m_{\text{r}}\,\frac{m^2}{\xi_+} \,\left(\frac{\partial \xi_{+}}{\partial x}\right)^2.
	\end{equation}
\end{linenomath*}
Finally, the difference between the energies of the two sides of the front acts as driving force for the front motion.
The kinetic energy of the front is then characterized by the mass of the front $m_{\text{rr}}\,\xi_{\text{eq}}$ and the front velocity $v_{+}$
\begin{linenomath*}
	\begin{equation}
		\label{Ek}
		E_{\text{kin}} = \frac{1}{2} \,m_{\text{rr}}\,\xi_{\text{eq}} \,v_+^2.
	\end{equation}
\end{linenomath*}
Here, $m_{\text{rr}}$ represents the mass of a receptor pair including the  cell receptor and  the virus receptor which are bonded to each other.
Combining Eq. \eqref{E-}, \eqref{E+} and \eqref{Ek} leads to the expression
\begin{linenomath*}
	\begin{equation}
		\label{ic2}
		-\xi_{\text{eq}}\,C_{\text{b}} + \xi_{\text{eq}}\,\text{ln}\left( \frac{\xi_{\text{eq}}}{\xi_0}\right) + \frac{1}{2}B\kappa^2 - \left[\xi_{+}\text{ln}\left(\frac{\xi_{+}}{\xi_0}\right) + \frac{1}{2}\frac{m_{\text{r}}}{k\,T}\,\frac{m^2}{\xi_+} \,\left(\frac{\partial \xi_{+}}{\partial x}\right)^2\right] = \frac{1}{2} \,\xi_{\text{eq}}\frac{m_{\text{rr}}}{k\,T}\,v_+^2,
	\end{equation}
\end{linenomath*}
which is the final form of the supplementary condition and closes the formulation of the moving boundary problem.

\section{Numerical implementation and results for 1D case and helical viruses}
\label{Numerical implementation and results for 1D case and helical viruses}

\subsection{Implementation}
\label{Numerical implementation and results for 1d case and helical viruses, Implementation}
In summary, the change of the receptor distribution is described by a system of differential equations consisting of \eqref{diff}, \eqref{ic1b} and \eqref{ic2}.
The finite difference method has been chosen for the solution of the underlying system of differential equations.
According to this approach, all derivatives are replaced by expressions dependent on discrete values of the function for the nodes of a chosen lattice.
Thus the differential equations are transformed into a system of algebraic equations.
An implicit scheme is used, with the following approximations for the derivatives
\begin{linenomath*}
	\begin{equation}
		\label{derivatives}
		\frac{\partial \xi}{\partial t} = \frac{\xi_{i}^{j+1} - \xi_{i}^{j}}{\Delta t},\quad\frac{\partial \xi}{\partial x} = \frac{\xi_{i}^{j+1}-\xi_{i+1}^{j+1}}{\Delta x},\quad\frac{\partial^2 \xi}{\partial x^2} = \frac{\xi_{i-1}^{j+1}-2\xi_{i}^{j+1}+\xi_{i+1}^{j+1}}{\Delta x^2}.
	\end{equation}
\end{linenomath*}
Here, subscript $i$ denotes the spatial position and superscript $j$ denotes the time.
The implementation of relationships \eqref{derivatives} into the system \eqref{diff}, \eqref{ic1b} and \eqref{ic2} leads to the following discretized formulation of the problem:
\begin{linenomath*}
	\begin{subequations}
		\label{equationsystem2}
		\begin{align}
			\label{equationsystem2a}
			&\hspace*{-2mm}\frac{\xi_{i}^{j+1} - \xi_{i}^{j}}{\Delta t} = m\frac{\xi_{i-1}^{j+1}-2\xi_{i}^{j+1}+\xi_{i+1}^{j+1}}{\Delta x^2},\quad i = 1,...,p,\quad j=1,...,n,\\
			\label{equationsystem2b}
			&\hspace*{-2mm}\left[\xi_{\text{eq}}-\xi_{+}^{j+1}\right]v_+^{j+1} + m\,\frac{\xi_{+}^{j+1}-\xi_{1}^{j+1}}{\Delta x} = 0,\\
			\label{equationsystem2c}
			&\hspace*{-2mm}\left[ E^- \right] - \left[\xi_{+}^{j+1}\text{ln}\left(\frac{\xi_{+}^{j+1}}{\xi_0}\right) + \frac{1}{2}\frac{m_{\text{r}}}{k\,T}\,\frac{m^2}{\xi_+^{j+1}} \,\left(\frac{\xi_{1}^{j+1}-\xi_{+}^{j+1}}{\Delta x}\right)^2\right] - \left[\frac{1}{2} \,\xi_{\text{eq}}\frac{m_{\text{rr}}}{k\,T} \,\left.v_+^{j+1}\right.^2\right] = 0.
		\end{align}
	\end{subequations}
\end{linenomath*}
In Eq. \eqref{equationsystem2a}, variable $p$ refers to the total number of points ahead of the front, except the last point where the influence of the flux vanishes and where the receptor density is kept at the initial value $\xi_0$.
Variable $n$ refers to the total number of time steps.
Furthermore, the conditions given in Eqs. \eqref{equationsystem2b}-\eqref{equationsystem2c} are valid at the front.
In Eq. \eqref{equationsystem2c}, $E^-$ is an abbreviation for the contribution defined in \eqref{E-}. This term does not depend on the density $\xi_{i}$ and quantities at the front $\xi_{+}$ and $v_+$, and thus represents a constant during the process.

The solution of system  	\eqref{equationsystem2} yields values  for receptor densities $\xi_{+}^{j+1}$ and $\xi_{i}^{j+1},$ $i = 1,...,p,$ and front velocity $v_{+}^{j+1}$. The latter is, in postprocessing, used to  evaluate  the front position  in the incremental form: $a^{j+1}=a^{j}+v_{+}^{j+1}\Delta\,t$, where $j+1$ and $j$ are two subsequent time steps.

\subsection{Results}
\label{Numerical implementation and results for 1D case and helical viruses, Results}

The numerical examples chosen simulate the process of virus uptake into the cell.
In the simulations, it is assumed that a helical virus of size $D = 0.05$~$\si{\micro}$m, comes into contact with a much larger cell such that the cell curvature is negligible (Fig. \ref{fig:P1_1}). 	Due to the axial symmetry of the virus, the problem is treated in a 1D representation which assumes the unit width of the active domain.
The initial density of cell receptors is set to $\xi_0 = 1000$~\si{\micro}m$^{-2}$, whereas the initial density of virus receptors is set to $\xi_{\text{eq}} = 4800$~\si{\micro}m$^{-2}$.
Time increment $\Delta t = 1e^{-4}$~s and space increment $\Delta x = 1e^{-3}$~\si{\micro}m are used for the numerical simulations. An overview of the chosen process parameters  is given in Table \ref{tab:P1_1}. These values belong to the corresponding admissible ranges and are also used in \cite{fre1}. The convergence of results has been checked by varying the time and the space increment. Time increment $\Delta t$ has been decreased in the interval $\Delta t = 1e^{-3} - 1e^{-5}$~s which has caused a change of results in $\xi_{+}$ for maximally $0.3\%$. The variation of the space increment $\Delta x$ in range $1e^{-2} - 1e^{-4}$~\si{\micro}m caused the changes in $\xi_{+}$ up to $2.5\%$. For both parameters, decreasing the increment by a constant factor reduces the error successively.
\begin{table}
	\centering
	\renewcommand*{\arraystretch}{1.2}
	\begin{tabular}{l l c r}
		\hline
		\bf Material parameters&&&\\
		\hline
		Receptor density on cell surface & $\xi_0$ & 1000 & \si{\micro}m$^{-2}$\\
		Receptor density on virus & $\xi_{\text{eq}}$ & 4800 & \si{\micro}m$^{-2}$\\
		Receptor mass & $m_{\text{r}}$ & 400 & kDa\\
		Mass of a receptor pair & $m_{\text{rr}}$ & 800 & kDa\\
		Binding energy coefficient& $C_{\text{b}}$ & 5 & $-$\\
		Numerical bending stiffness parameter & $B$ & 30 & $-$\\
		Curvature of the virus & $\kappa$ & $40$ & $\si{\micro}\text{m}^{-1}$\\
		Mobility parameter & $m$ & $0.5$-$1$ & $\si{\micro} \text{m}^2/$s\\
		Virus diameter & $D$ & $0.05$ & $\si{\micro} \text{m}$\\
		\hline
	\end{tabular}
	\caption{Process parameters used in simulations.}
	\label{tab:P1_1}
\end{table}

\begin{figure}[H]
	\centering
	\includegraphics[width = 1.\textwidth]{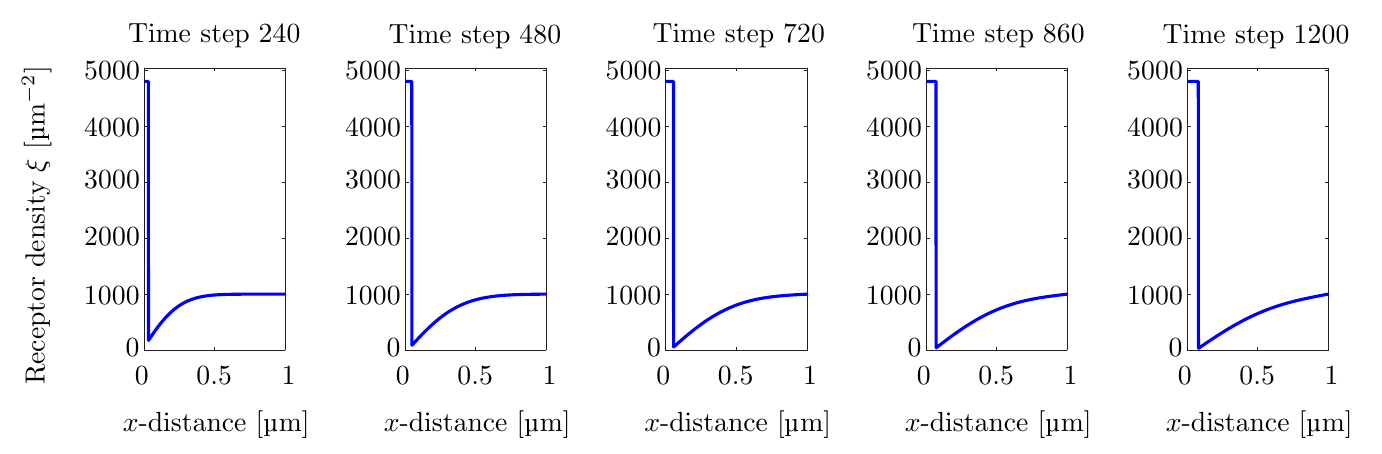}
	\caption{Receptor density $\xi$ over the cell surface $x$ for the first $1200$ time steps, corresponding to a simulated time of $0.12\,$s.}
	\label{fig:P1_3}
\end{figure}

The first group of simulations studies the change of the cell receptor density during the process and the front motion for the mobility parameter set to $m = 1$~\si{\micro}m$^{2}$/s.
The density profiles for different time steps during the simulation are presented in Fig. \ref{fig:P1_3}. The diagrams show a fast decrease in receptor density,  particularely at the beginning of the process. After 300 time steps, the density at the front only amounts to $\approx 50 \%$ of its initial value. This rapid decline in density at the front slows down in the course of the further process.

\begin{figure}[H]
	\centering
	\includegraphics[width = 1.\textwidth]{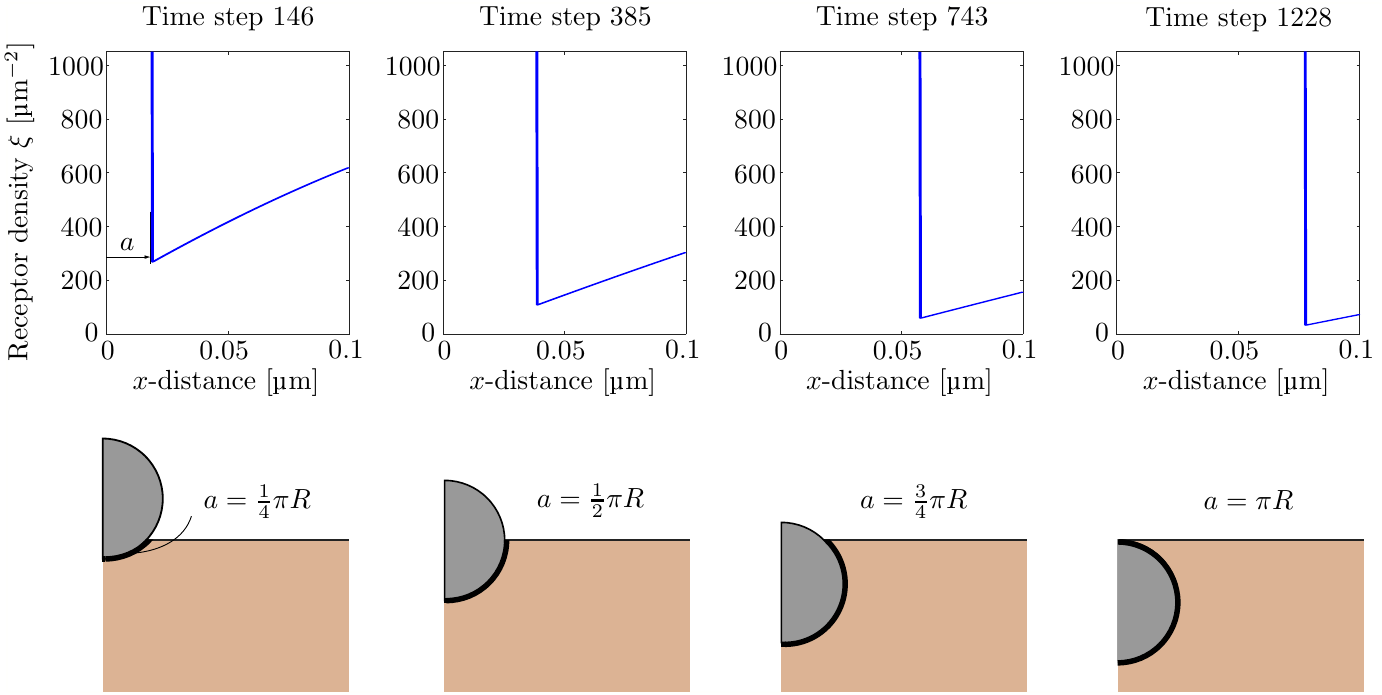}
	\caption{Visualization of the front motion and of the formation of the envelope around the virus with diameter $D = 0.05~\si{\micro}$m.}
	\label{fig:P1_4}
\end{figure}

\begin{figure}
\centering
\includegraphics[width = 1.\textwidth]{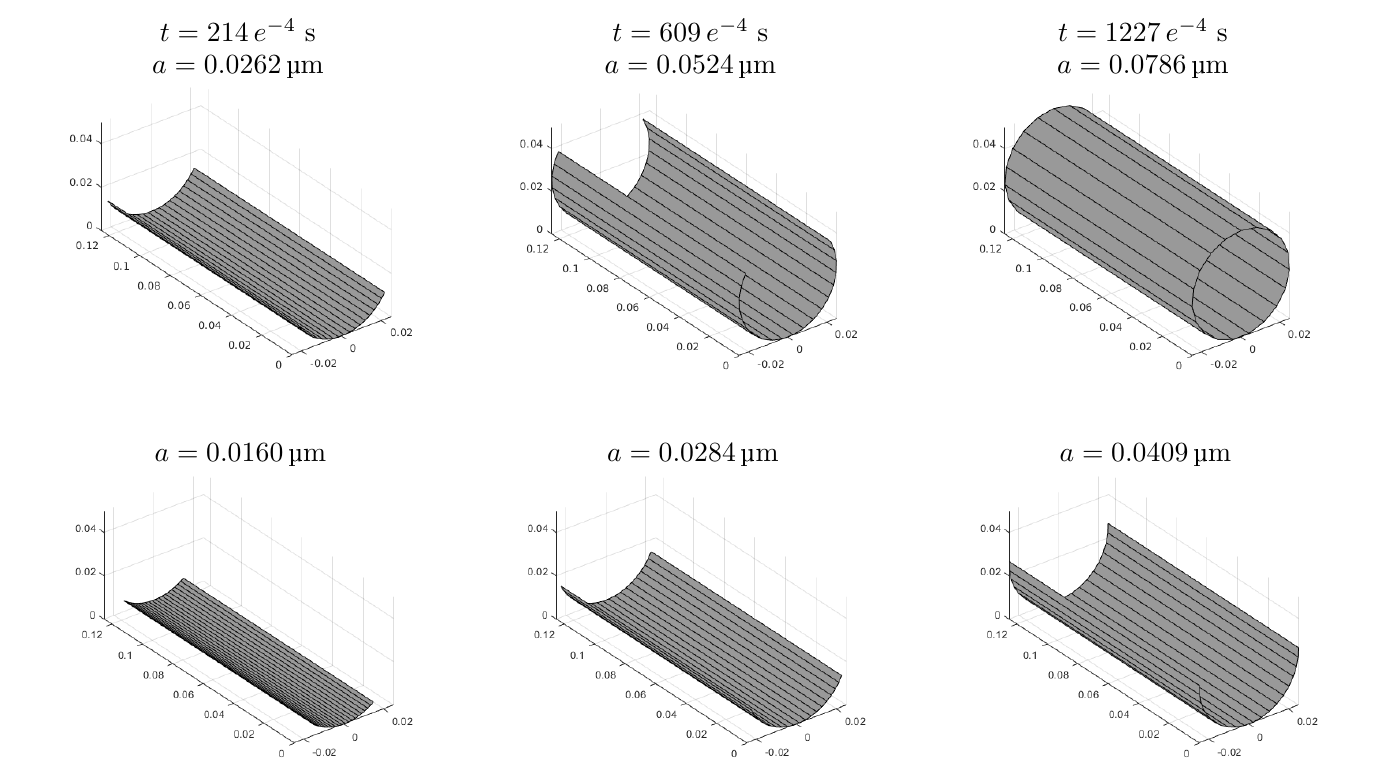}
\caption{Endocytosis of a helical virus. Comparison of the process state at the same time for two different mobilities. The upper row shows results corresponding to  a higher mobility $m = 1\,\si{\micro} \text{m}^2/$s. Results presented in the bottom row are related to lower mobility $m = 0.5\,\si{\micro} \text{m}^2/$s.}
\label{fig:P1_13}
\end{figure}

Figure \ref{fig:P1_4} monitors the advancement of the front and the position of the virus during its entry into the cell in 1D representation. The position of the virus is related to the position of the front through length $a$, determining the size of contact area.
For a helical virus,  a 3D visualization is also possible due to the axial symmetry, as shown in Fig.~\ref{fig:P1_13} which compares the endocytosis of a virus into cells with different receptor mobilities. In the top row,  the mobility is set to $m = 1\,\si{\micro} \text{m}^2$/s, whereas half of this value  $m = 0.5\,\si{\micro} \text{m}^2$/s is used in simulations  in the lower row. Naturally, the first process  is faster and the viral entry is accomplished  earlier than in the second case.
The increasing number of time steps between the three states indicates the gradual decrease and final stagnation of the velocity of the process, an issue also studied in the following example.

\begin{figure}[H]
	\centering
	\includegraphics[width = 1.\textwidth]{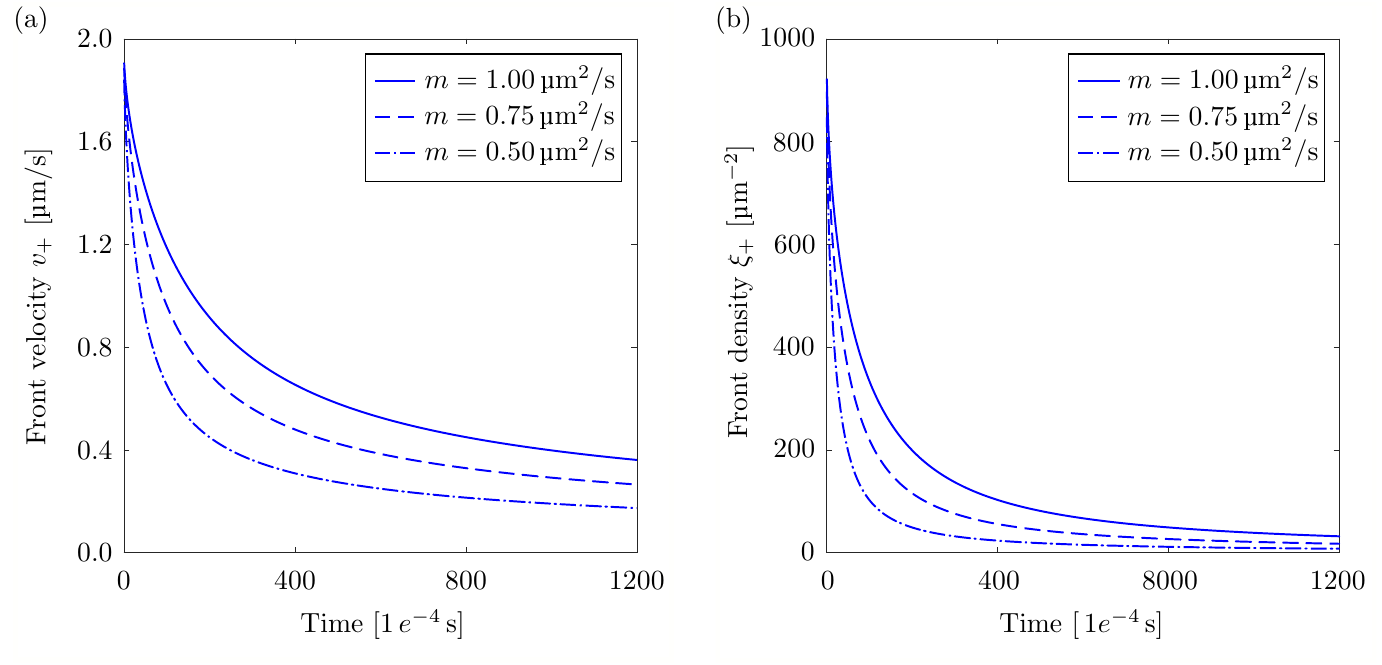}
	\caption{(a) Velocity of the adhesion front vs. time. (b) Evolution of the receptor density at the adhesion front. Mobility is varied in the range $0.5\,\si{\micro}\text{m}^2/\text{s}-1\,\si{\micro}\text{m}^2/\text{s}$.}
	\label{fig:P1_6}
\end{figure}

The governing equation \eqref{diff} of the process depends on a single process parameter, namely on mobility $m$. The parameter represents a measure for the capability of receptors to move over the cell surface, and thus is in a direct correlation with the amount of receptors provided for the adhesion with the virus. The influence of the mobility on the velocity of the front and on the receptor density has been studied on the basis of a set of simulations, as shown in Fig. \ref{fig:P1_6}. Here, mobility parameter $m$ has been varied in the range $[0.5~\si{\micro}\text{m}^2/\text{s}$-$1~\si{\micro}\text{m}^2/\text{s}]$. Figure \ref{fig:P1_6}a shows the dependence of velocity $v_+$ on the mobility and clearly confirms the rapid decrease in velocity at the beginning followed by a stagnation, as already observed in the previous test (Fig. \ref{fig:P1_4}). The value of the mobility does not affect the form of the velocity diagrams. However,  a higher velocity corresponds to a higher mobility. This observation is in agreement with the physical character of the mobility describing the ability of receptors to move towards the adhesion zone. For lower values of $m$, fewer receptors are provided to connect the cell with the virus. Therefore, the evolution of the adhesion zone and the velocity of the front are slowed down. An analogous trend is observed for the dependency of the receptor density at the front on the mobility shown in Fig. \ref{fig:P1_6}b.

\begin{figure}
	\centering
	\includegraphics[width = 1.\textwidth]{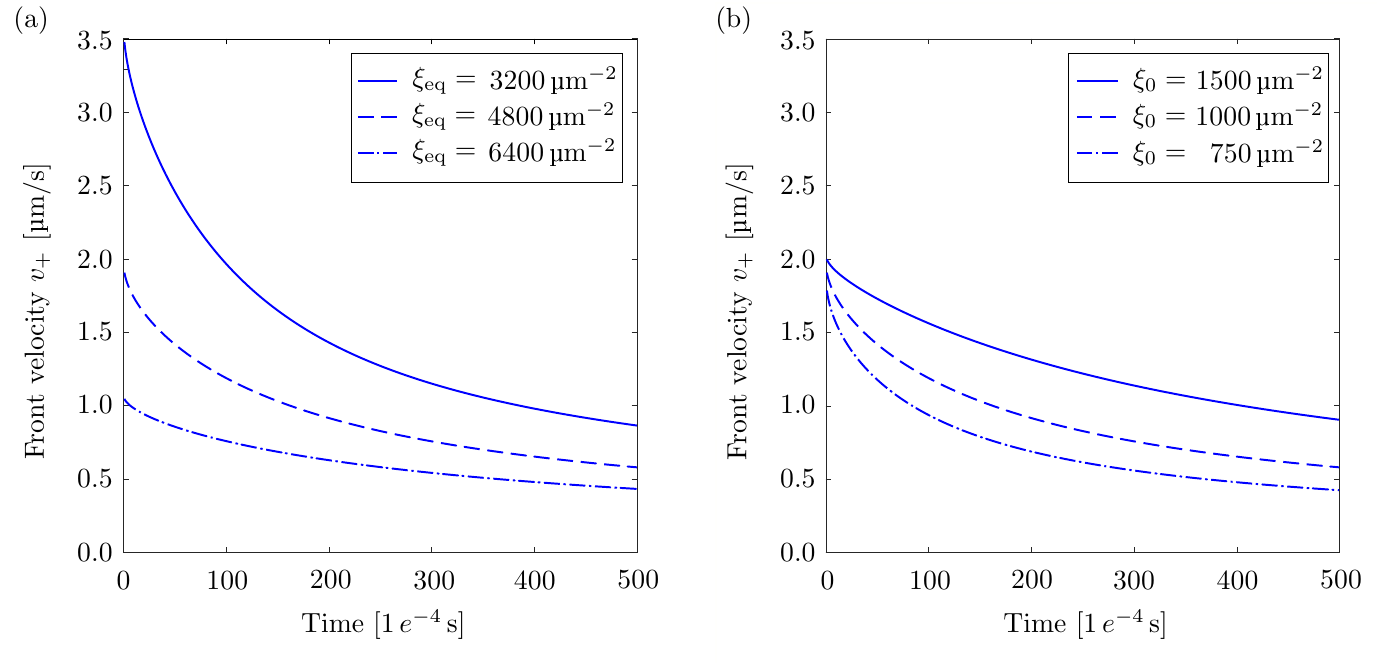}
	\caption{(a) Evolution of the velocity of the adhesion front for different densities $\xi_{\text{eq}}$. Density $\xi_0$ is set to $1000$ \si{\micro}m$^{-2}$. (b) Evolution of the velocity at the adhesion front front for different initial densities $\xi_0$. Density $\xi_{\text{eq}}$ is set to $4800$ \si{\micro}m$^{ -2}$.}
	\label{fig:P1_7}
\end{figure}

An important influence on the process is also imposed by the fixed receptor density $\xi_{\text{eq}}$ of the virus, initially chosen to dictate the amount of receptors required for the virus-cell connection.
The velocity of the adhesion front $v_+$ for different values of $\xi_{\text{eq}}$ is shown in Fig. \ref{fig:P1_7}a.
Here, the receptor density of the cell is set to $\xi_0 = 1000$~\si{\micro}m$^{-2}$ and the mobility is set to $m = 1$~\si{\micro}m$^{2}$/s.
The form of the velocity diagrams does not change, although the different constellations are taken into consideration.
The velocity of the adhesion front $v_+$ decreases with increasing density $\xi_{\text{eq}}$, which is to be expected since a larger number of receptors is necessary in order to achieve a front advancement.

Similar simulations are conducted for different values of the initial receptor density $\xi_0$, while the receptor density of the virus is set to $\xi_{\text{eq}} = 4800$~\si{\micro}m$^{-2}$ (Fig. \ref{fig:P1_7}b).
Again, the initial configuration does not affect the form of the diagrams, whereas a larger density $\xi_0$ corresponds to higher velocities.
The required amount of connected receptors has been fixed at a constant value in all the simulations.
However, only the initiation of the process requires a higher amount of bonds.
Once contact between the cell and the virus has been established, the number of necessary receptors decreases.
The amount of bonds required for the contact between the cell and the virus cannot fall below a minimum value.
The evolution of the required cell receptor density can be easily implemented in the developed code by assuming $\xi_{\text{eq}}$ to be a function of time.
The simulations in this case (results not shown here) indicate an accelerated viral entry into the cell as a consequence of the decrease of the required receptor density.


\section{Non-dimensionalization}
\label{Nondimensionalization}

\subsection{Derivation of the non-dimensional formulation}
\label{Nondimensionalization, Derivation}
In some physical systems, non-dimensionalization is applied to suggest that it is more convenient to measure  certain quantities relative to an appropriate unit. These units refer to quantities intrinsic to the system. The non-dimensionalization procedure relies on replacing dimensional quantities by the non-dimensional ones within the differential equation and within the corresponding boundary and supplementary conditions. An important advantage of the non-dimensional analysis is that it reduces the number of relevant process parameters and thus facilitates the parameter study as performed in the previous section.

The present model for the viral entry includes following dimensional quantities which can be expressed in terms of their non-dimensional counterparts
\begin{linenomath*}
	\begin{equation}\label{variab}
		x = l_{\text{s}} \,x', \qquad	t = \tau_{\text{s}}\, t', \qquad \xi=\xi_{\text{s}} \,\xi'.
	\end{equation}
\end{linenomath*}
Here,  $l_{\text{s}}$, $\tau_{\text{s}}$, $\xi_{\text{s}}$ are properly chosen scaling parameters for space, time and density,  and the prime symbol denotes the dimensionless quantities.
The introduction of transformation \eqref{variab} in differential equation \eqref{diff} yields its non-dimensional form
\begin{linenomath*}
	\begin{equation}
		\frac{\partial \xi'}{\partial t'} = \frac{m\,\tau_{\text{s}}}{l_{\text{s}}^2}\frac{\partial^2 \xi'}{\partial x'^{\,2}}.
		\label{ndim1}
	\end{equation}
\end{linenomath*}
The same procedure can now be applied to the flux-boundary condition \eqref{ic1b} and the supplementary energy condition \eqref{ic2} which, amongst others,  depend on front velocity $v_+$. The latter intrinsically includes  the derivative with respect  to time, which yields
\begin{linenomath*}
	\begin{equation} \label{velocity}
		v_+ = \frac{\partial a}{\partial t} = \frac{\partial (l_{\text{s}}\,a')}{\partial (\tau_{\text{s}}\,t')} =\frac{l_{\text{s}}}{\tau_{\text{s}}} v_+'.
	\end{equation}
\end{linenomath*}
Bearing in mind transformations \eqref{variab} and intermediate result \eqref{velocity}, the flux boundary condition and the energetic  supplementary condition have the non-dimensional form
\begin{linenomath*}
	\begin{gather}
		(\xi'_{\text{eq}}-\xi'_+)\,v_+' - \frac{m\,\tau_{\text{s}}}{l_{\text{s}}^2}\left[ \frac{\partial \xi'}{\partial x'} \right]_+ = 0,
		\label{ndim2a}
		\\
		C'_1 - \xi'_{+}\,\text{ln}\,\xi'_{+} - \frac{1}{2}\,\frac{m_{\text{r}}}{k\,T}\,\frac{m^2}{l_{\text{s}}^2}\,\frac{1}{\xi'_+}\left(\frac{\partial \xi'_{+}}{\partial x'}\right)^2 = \frac{1}{2}\,\frac{m_{\text{rr}}}{k\,T}\, \frac{l_{\text{s}}^2}{\tau_{\text{s}}^2}\, \xi'_{\text{eq}}\,v_+'^2\, ,
		\label{ndim3}
	\end{gather}
\end{linenomath*}
where the following abbreviations apply
\begin{linenomath*}
	\begin{gather}
		\xi'_{+}=\frac{\xi_+}{\xi_{\text{s}}},\qquad\xi'_{\text{eq}}=\frac{\xi_{\text{eq}}}{\xi_{\text{s}}},\\
		C'_1 = 	-\xi'_{\text{eq}}\,C_{\text{b}} + \xi'_{\text{eq}}\,\text{ln}\,\xi'_{\text{eq}} + \frac{1}{2}B\,\frac{\kappa^2}{\xi_{\text{s}}}.
	\end{gather}
\end{linenomath*}
Previous formulation \eqref{ndim2a}-\eqref{ndim3} calls upon the  introduction of additional non-dimensional process parameters
\begin{linenomath*}
	\begin{equation}
		\overline{m} = \frac{m\,\tau_{\text{s}}}{l_{\text{s}}^2},\qquad \overline{m}_{\text{r}} = \frac{1}{2}\frac{m_{\text{r}}}{k\,T}\,\frac{m^2}{l_{\text{s}}^2},\qquad \overline{m}_{\text{rr}} = \frac{1}{2}\frac{m_{\text{rr}}}{k\,T}\, \frac{l_{\text{s}}^2}{\tau_{\text{s}}^2}\,\xi'_{\text{eq}}.
	\end{equation}
\end{linenomath*}
where $\overline{m}$ represents the dimensionless mobility, $\overline{m}_{\text{r}}$ is the non-dimensional cell receptor mass and $\overline{m}_{\text{rr}}$ the non-dimensional mass of a receptor couple. This short notation yields the final form of the non-dimensional moving boundary problem
\begin{linenomath*}
	\begin{gather}
		\frac{\partial \xi'}{\partial t'} = \overline{m}\,\,\frac{\partial^2 \xi'}{\partial x'^{\,2}},\\[1mm]
		(\xi'_{\text{eq}}-\xi'_+)\,v_+' - \overline{m}\left[ \frac{\partial \xi'}{\partial x'} \right]_+ = 0,
		\label{ndim2}
		\\
		C'_1 - \xi'_{+}\,\text{ln}\,\xi'_{+} - \overline{m}_{\text{r}}\,\frac{1}{\xi'_+}\left[\frac{\partial \xi'_{+}}{\partial x'}\right]^2 = \overline{m}_{\text{rr}}\,v_+'^2\, .
	\end{gather}
\end{linenomath*}

\subsection{Analysis and results}
\label{Nondimensionalization, Analysis and results}
After deriving the non-dimensional formulation of the problem, a further important step  is choosing characteristic scaling parameters. Typically, these are adapted  to the system properties. In the present case, half of the arclength of the virus is assumed as the characteristic length, namely $l_{\text{s}}=\pi R$. On the other hand, the characteristic time is chosen as the time  necessary  to complete the virus uptake at a constant unit velocity $v_0=1\frac{\si{\micro}\text{m}}{\text{s}}$. Accordingly,  the time scaling parameter reads $\tau_{\text{s}}=l_{\text{s}}/v_0$. The initial cell receptor density is chosen as the last scaling parameter, such that it holds $\xi_{\text{s}}=\xi_0$.

The results for a non-dimensional analysis are presented by examples investigating the influence of non-dimensional parameters $\overline{m},\,\,\overline{m}_{\text{r}}$, $\overline{m}_{\text{rr}}$, $C'_1$ and $\xi'_{\text{eq}}$ on the front velocity and front receptor density. To this end, first,  the reference values for the scaling  parameters are set  as follows:  $ l_{\text{s}}=\pi R=0.0785\,\,\si{\micro}\text{m}$ and $\tau_{\text{s}}=l_{\text{s}}/v_0 =0.0785\,\,\text{s}$. Here, radius $R=0.025\,\, \si{\micro}\text{m}$  is assumed as the reference virus size. The density scaling parameter takes the value  $\xi_{\text{s}}=\xi_0= 1000\,\si{\micro}\text{m}^{-2}$.

\begin{figure}
	\centering
	\includegraphics[width = 0.95\textwidth]{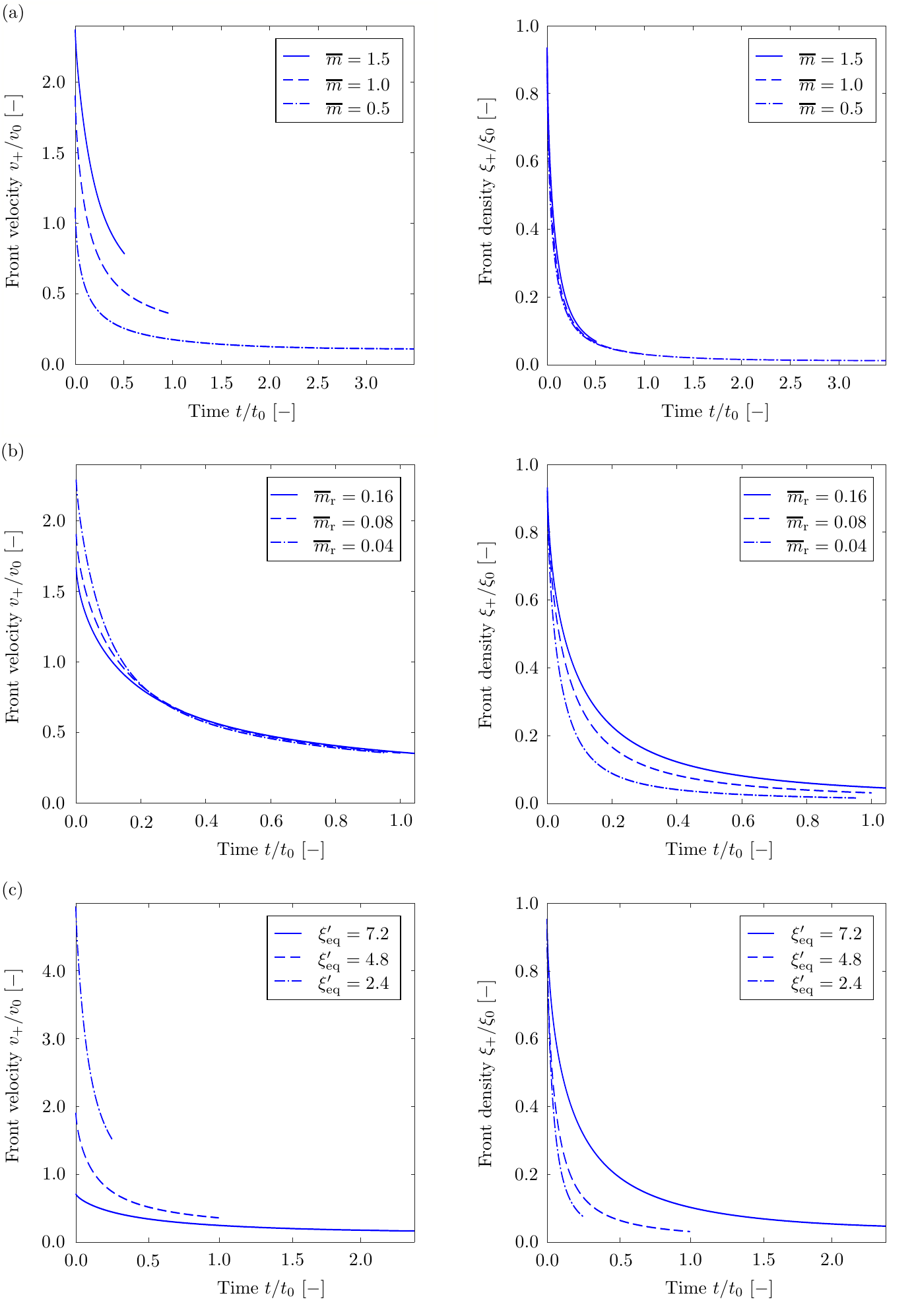}
	\caption{The non-dimensional analysis of the front velocity and the front density. (a) The influence of  non-dimensional mobility $\overline{m}$. Remaining parameters are kept constant: $\overline{m}_\text{r}= 0.08$, $\overline{m}_{\text{rr}}= 0.78$, $C'_1 = 7.53$, $\xi'_\text{eq} = 4.8$. (b) The influence of  non-dimensional receptor mass $\overline{m}_\text{r}$. Remaining parameters: $\overline{m}= 1.00$, $\overline{m}_{\text{rr}}= 0.78$, $C'_1 = 7.53$, $\xi'_\text{eq} = 4.8$. (c) The influence of non-dimensional equilibrium density $\xi'_\text{eq}$. Remaining parameters: $\overline{m}= 1.00$, $\overline{m}_{\text{r}}= 0.08$, $\overline{m}_{\text{rr}}= 0.78$, $C'_1 = 7.53$.
	}
	\label{fig:P1_12}
\end{figure}

The results of the analysis parameter analysis are presented in Fig.~\ref{fig:P1_12}.
They show that  the dimensionless mobility $\overline{m}$ has an important  influence on the  front velocity (Fig.~\ref{fig:P1_12}a), whereas the dimensionless mass $\overline{m}_{\text{r}}$ mainly influences  the front receptor density (Fig.~\ref{fig:P1_12}b). The variation of the dimensionless equilibrium density $\xi'_{\text{eq}}$ has an important influence on both quantities
(Fig.~\ref{fig:P1_12}c). The effect of $C'_1$ is similar to the one of $\xi'_{\text{eq}}$, whereas the variation of the dimensionless couple mass $\overline{m}_{\text{rr}}$ hardly affects the results (results not shown here). Amongst others, a higher mobility corresponds  to the higher front velocity, whereas higher mass $\overline{m}_{\text{r}}$  and equilibrium density $\xi'_{\text{eq}}$ cause a higher front density.


\section{Rotationally symmetric case - Spherical virus}
\label{Rotationally symmetric case - Spherical virus}

\begin{figure}
	\centering
	\includegraphics[width = 1.\textwidth]{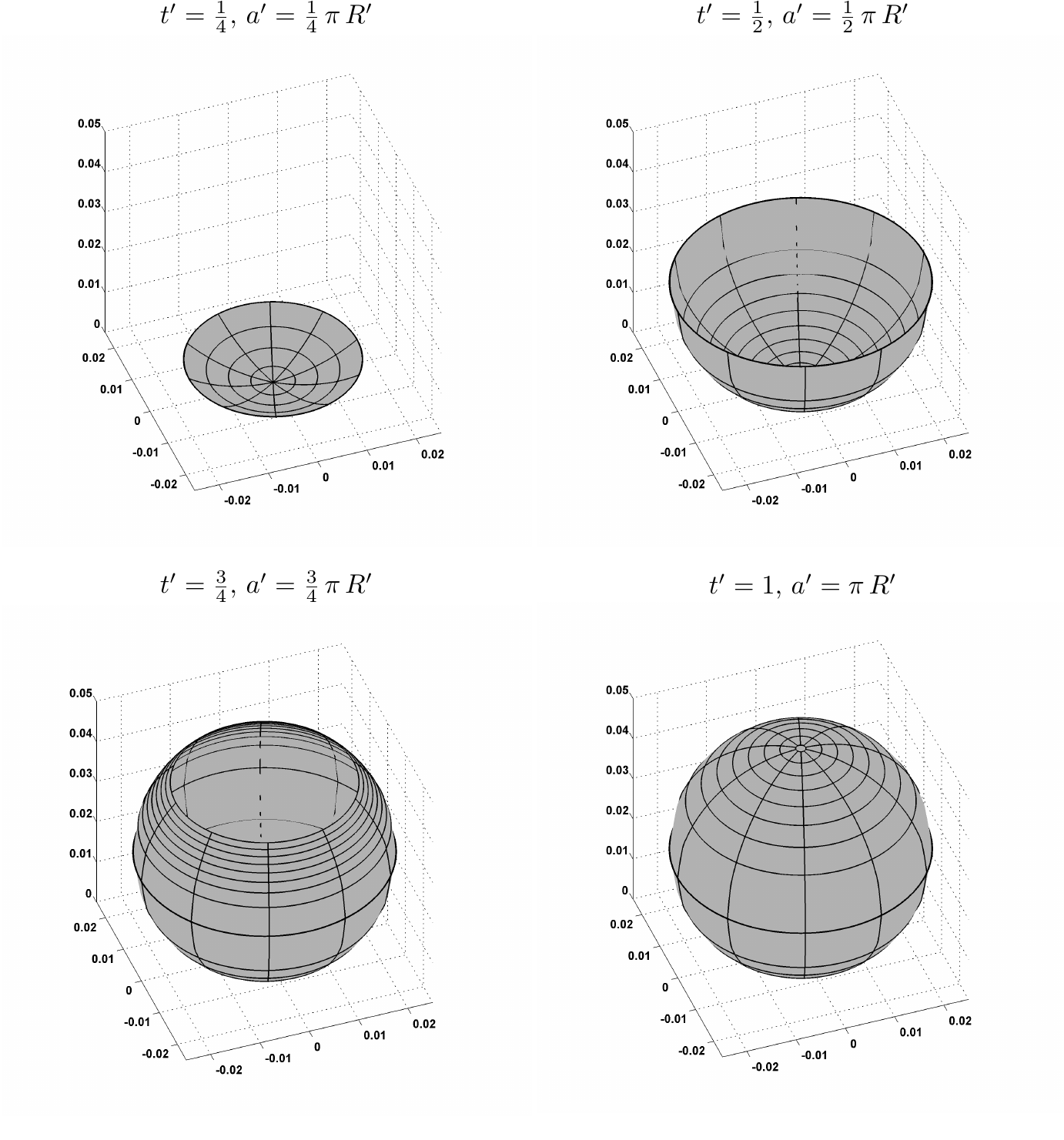}
	\caption{The  3D visualization of the endocytosis of a spherical virus with diameter $D = 0.05~\si{\micro}$m.}
	\label{fig:P1_5}
\end{figure}

Whereas the previous analysis focuses on the 1D formulation applicable for the simulation of  helical viruses, the real situation is commonly concerned with the rotationally symmetric geometry and spherical viruses. The problem formulation in this case is slightly different and  requires an adaptation of the diffusion equation, whereas the  boundary  and supplementary conditions remain unchanged. The extension of the diffusion equation implies the introduction of an additional term compensating for the radial dependency. Thus, the differential equation turns into
\begin{linenomath*}
	\begin{equation}
		\frac{\partial \xi}{\partial t} - m\frac{\partial^2 \xi}{\partial x^2} - m\frac{1}{x}\frac{\partial\xi}{\partial x} = 0,
		\label{diff3d}
	\end{equation}
\end{linenomath*}
where the last term on the left-hand side is the new contribution.
The discretized counterpart of Eq.~\eqref{diff3d} is
\begin{linenomath*}
	\begin{equation}\label{sphdis}
		\frac{\xi_{i}^{j+1} - \xi_{i}^{j}}{\Delta t} - m\frac{\xi_{i-1}^{j+1}-2\xi_{i}^{j+1}+\xi_{i+1}^{j+1}}{\Delta x^2} - m\,\frac{1}{x}\,\frac{\xi_{i+1}^{j+1} - \xi_{i-1}^{j+1}}{2\,\Delta x} = 0,
	\end{equation}
\end{linenomath*}
where $i = 1,...,p$ is the counter related to the  spacial discretization and where $ j=1,...,n$ is the counter  corresponding to the  time discretization. Variable $x$ in the last term in \eqref{sphdis} represents the distance from  the first contact point and also can be written in a discretized form as $x=i\Delta \,x$, which leads  to a condensed discretization formulation
\begin{linenomath*}
	\begin{equation}
		\frac{\xi_{i}^{j+1} - \xi_{i}^{j}}{\Delta t} - \frac{m}{2i\,\Delta x^2}
		\left[\left(2i-1\right)\xi_{i-1}^{j+1}-4i\xi_{i}^{j+1}+\left(2i+1\right)\xi_{i+1}^{j+1}\right]=0 \,.
		\label{diff3ddiscretized}
	\end{equation}
\end{linenomath*}
Finally,  the procedure explained in Sect. \ref{Nondimensionalization, Derivation}  provides the non-dimensinonal form of Eq.~\eqref{diff3d}:
\begin{linenomath*}
	\begin{equation}
		\frac{\partial \xi'}{\partial t'} - \overline{m}\,\frac{\partial^2 \xi'}{\partial x'^{\,2}} - \overline{m}\,\frac{1}{x'}\frac{\partial\xi'}{\partial x'}=0 \,.
	\end{equation}
\end{linenomath*}
The simulation of the  virus uptake for a rotationally symmetric case is demonstrated  by the example of the  Alphavirus and by the process parameters  summarized in Tab. \ref{tab:P1_1}. The achieved results are  summarized in Fig.~\ref{fig:P1_5}.  They show the 3D vesicle that is built during the process of the viral entry. Different to the endocytosis of a helical virus, the process advances at a rather constant rate throughout the simulation. This change in behavior can be explained by providing additional receptors due to the radial dependence. The same argumentation explains the higher velocity of the uptake process by a spherical virus compared to the helical one. For the chosen parameters, the simulations predict a required time in the range for ultra-fast-endocytosis \cite{watanabe2013}. This fast behavior is expected, since the mobility has a rather high value. Viruses often connect to receptors with a lower mobility. By reducing mobility to $m=0.2\,\,\si{\micro}\text{m}^2/\text{s}$ the time increases and matches values  for  kiss-and-run-endocytosis  \cite{zhang2009}.  The model predicts a shorter duration of the process than it is typical of the clathrin-mediated-endocytosis (15-20 s) \cite{balaji2007}.


\section{Cooperativity}
\label{Cooperativity}
\begin{figure}
	\centering
	\includegraphics[width = 1.\textwidth]{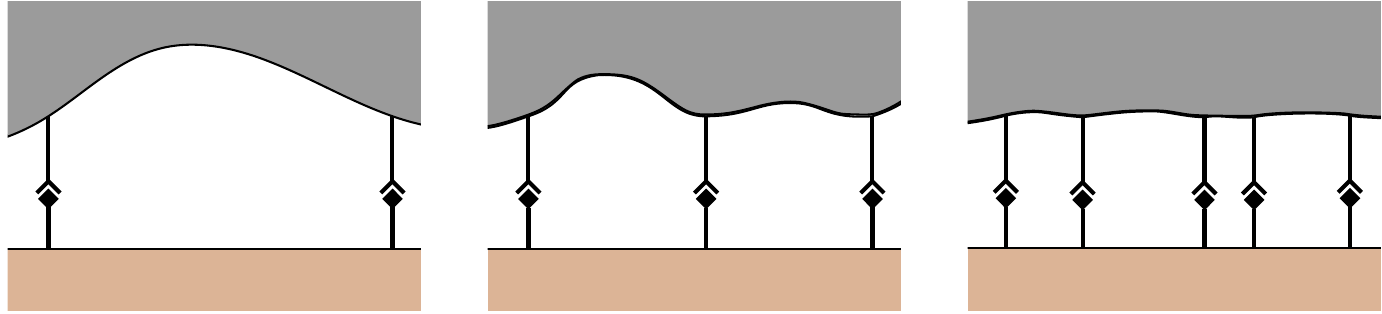}
	\caption{Schematic representation of cooperativity during endocytosis (the free receptors are not shown).}
	\label{fig:P1_8}
\end{figure}

Amongst others, cell adhesion deals with cooperativity, an effect which is explained by considering a patch of unit length depicted in Fig. \ref{fig:P1_8}.
As soon as receptors create bonds, they smoothen out the surrounding membrane which makes it easier for additional receptors to create a bond and strengthens the adhesion between the virus and the cell membrane \cite{Krobath2009,Weikl:2009}.
This effect is known as cooperativity.
It has extensively been investigated experimentally and theoretically.
Different experiments are performed depending on the state of the adhesion process.
The fluorescence recovery experiments are performed in order to analyze the equilibrated contact zone during the process, whereas the micropipette experiments are performed in order to analyze the initial contact.
Lipid vesicles with anchored receptor molecules are often used in order to resemble important aspects of cell adhesion.

In order to study the binding cooperativity, two classes of numerical models are considered.
The first class describes the membranes as continuous in space with continuous concentration profiles on the membrane \cite{Wu:2006}.
The second class describes the membranes as discrete and the receptors as single molecules \cite{Krobath:2007}.
Numerical solutions of the dynamic properties are studied by reaction-diffusion equations in the first class \cite{Shenoy:2005} of models and by Monte Carlo simulations in the second class \cite{Tsourkas:2007}.
The information obtained in such a way is complementary to the model presented in this contribution.

The cooperativity changes the amount of receptor bonds that will create an equilibrium state upon connection between the virus and the cell given in \cite{Krobath2009} according to
\begin{linenomath*}
	\begin{equation}
		\label{xireq}
		\xi_{\text{eq\_req}} = c\frac{\kappa_\text{b}}{k\,T}l^2_{\text{we}}K^2_{\text{pl}}\xi_{\text{eq}}^2\xi_0^2.
	\end{equation}
\end{linenomath*}
Here, $\xi_{\text{eq\_req}}$ is the required amount of receptors that need to bind in order to create adhesion between the virus and the cell.
Symbol $c$ denotes a dimensionless prefactor acquired from Monte Carlo simulations, usually ranging between $10$ - $15$.
Furthermore, the effective rigidity $\kappa_\text{b}$ can be calculated from the bending rigidities of two apposing membranes as $\kappa_\text{b} = \kappa_1\,\kappa_2/(\kappa_1 + \kappa_2)$.
For the simulations here, it is set to $40\,k\,T$.
Quantity $l_{\text{we}}$ is the binding range depending on the interaction range of the two binding sites, of the flexibility of their molecules and of the membrane anchoring.
It describes the difference between the smallest and the largest local membrane separation at which the receptors can bind.
Quantity $K_{\text{pl}}$ is the two-dimensional equilibrium constant in the case of two opposing planar, supported membranes within binding separation of the receptor–ligand bonds.

\begin{figure}
	\centering
	\includegraphics[width = 1.\textwidth]{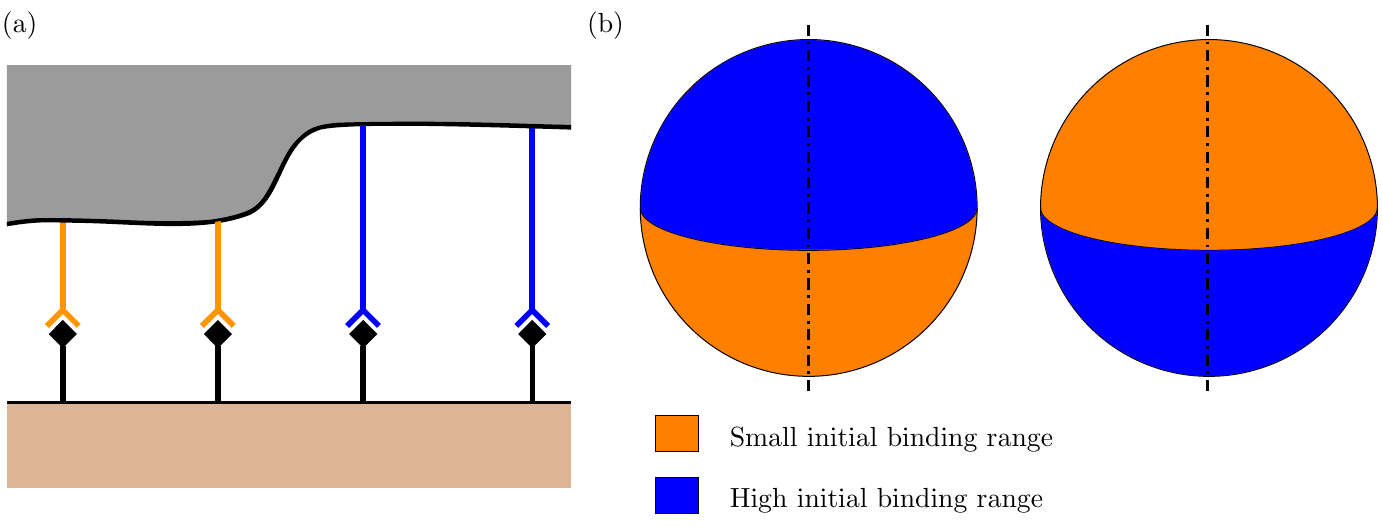}
	\caption{(a) Contact with the virus with two kinds of receptors. (b) Spatial distribution of different types of receptors on the virus membrane for two chosen configurations.}
	\label{fig:P1_9}
\end{figure}

Two illustrative examples are performed in order to analyze the influence of cooperativity for different binding ranges.
As in the previous example the initial receptor density of the cell is set to $\xi_{0} = 1000$~\si{\micro}m$^{-2}$ and the receptor density of the virus is set to $\xi_{\text{eq}} = 4800$~\si{\micro}m$^{-2}$.
The first group of simulations considers a virus with its lower half covered by receptors with a smaller binding range and its upper half by receptors with a larger binding range.
The lower half is characterized by a binding range of $l_{\text{we}} = 1$~nm resulting in the required receptor density $\xi_{\text{eq\_req}} = 2265$~\si{\micro}m$^{-2}$, while the upper half is characterized by a binding range of $l_{\text{we}} = 1.2$~nm resulting in the required receptor density $\xi_{\text{eq\_req}} = 3262$~\si{\micro}m$^{-2}$.
An opposite situation is considered in the second group of simulations presented in Fig. \ref{fig:P1_9}.

\begin{figure}
	\centering
	\includegraphics[width = \textwidth]{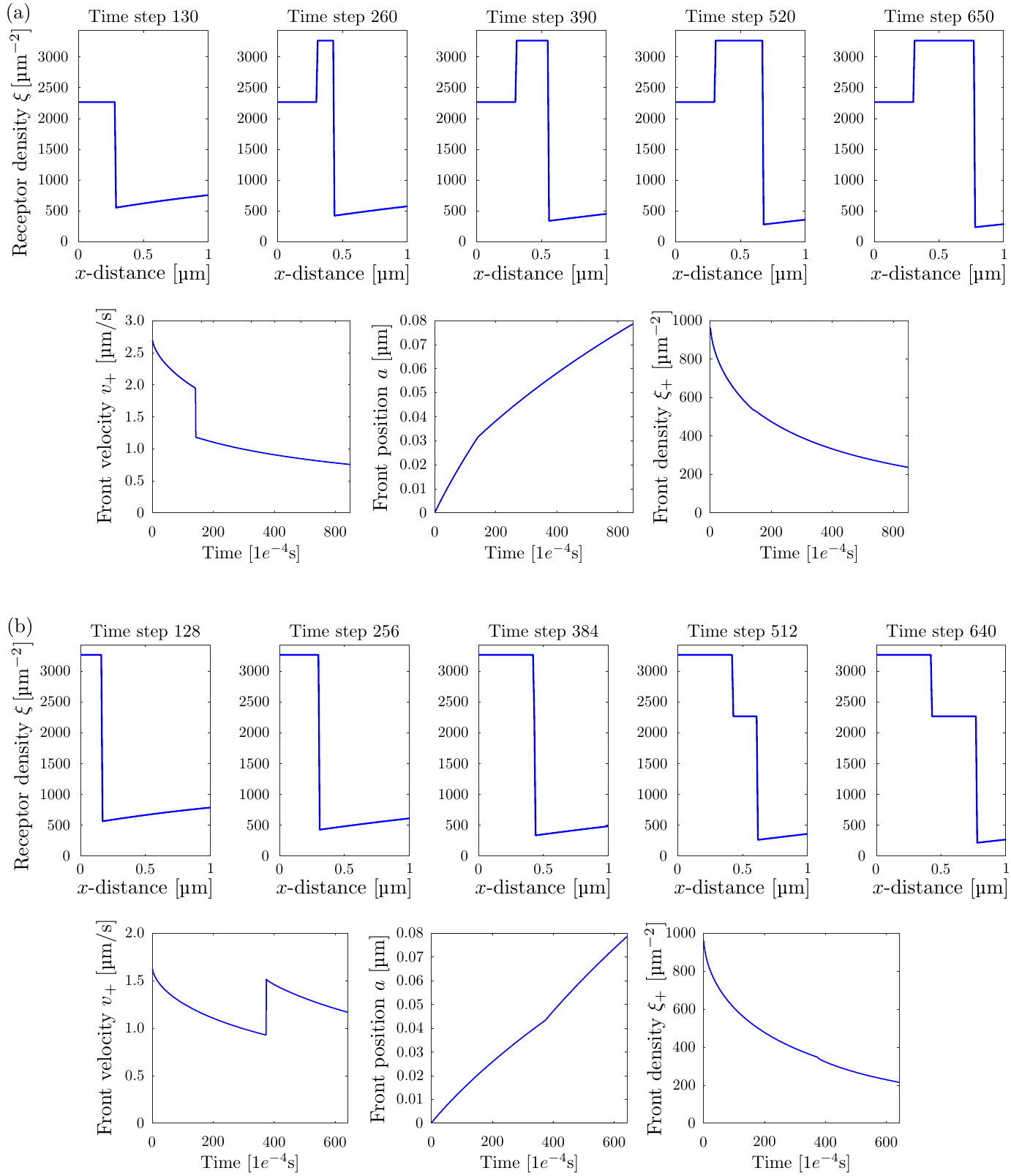}
	\caption{(a) Influence of the cooperativity for a smaller binding range in the lower half $\xi_{\text{eq\_req}} = 2265$ \si{\micro}m$^{-2}$ and a larger one in the upper half $\xi_{\text{eq\_req}} = 3262$ \si{\micro}m$^{-2}$.
		(b) Influence of the cooperativity for a larger binding range in the lower half $\xi_{\text{eq\_req}} = 3262$ \si{\micro}m$^{-2}$ and a smaller one in the upper half $\xi_{\text{eq\_req}} = 2265$ \si{\micro}m$^{-2}$. Top row: Receptor density over the cell surface for different time steps. Bottom row: Velocity of the front, position of the front and receptor density at the front over time.
		Chosen process parameters are $K_{\text{pl}} = 0.55\,e^{-3}$ and $c = 13$.}
	\label{fig:P1_10}
\end{figure}

Numerical results for the described examples are shown in Fig. \ref{fig:P1_10}.
The transition between the areas with different receptor types manifests itself by either a jump or a kink in the corresponding diagrams.
The velocity is affected mostly by the change of the required density.
In the area with a smaller binding range less receptors are required, significantly increasing the velocity of the process.
The diagrams for the second setup show similar results to the first setup, however, the change from the lower to the upper half is significantly delayed.
Here, the initial velocity is much higher in the first case such that the virus is almost enclosed at the time step 400.
Contrary to this, the velocity at the end of the process is higher in the second case.
Consequently, both viruses need approximately 600 time steps for their entry into the cell.
Exact values are 611 and 622 time steps for the first and second example respectively.
The values do not exactly match due to the different velocities at the beginning of the process and due to the transition between regions with different receptors.

\section{Discussion}
\label{Discussion}

The model developed  gives insight into some specific features of the process and enables its profound analysis in the context of impeding and  hindering the viral entry. Amongst others, it enables a study of the position of the  front and its velocity during the process, an analysis of the admissible values for the radius and of the duration of the process depending on different process parameters. The study presented uses the parameters listed in Tab. \ref{tab:P1_1}, if not stated otherwise.

\subsection{Front position and velocity}
\begin{figure}
	\centering
	\includegraphics[width = \textwidth]{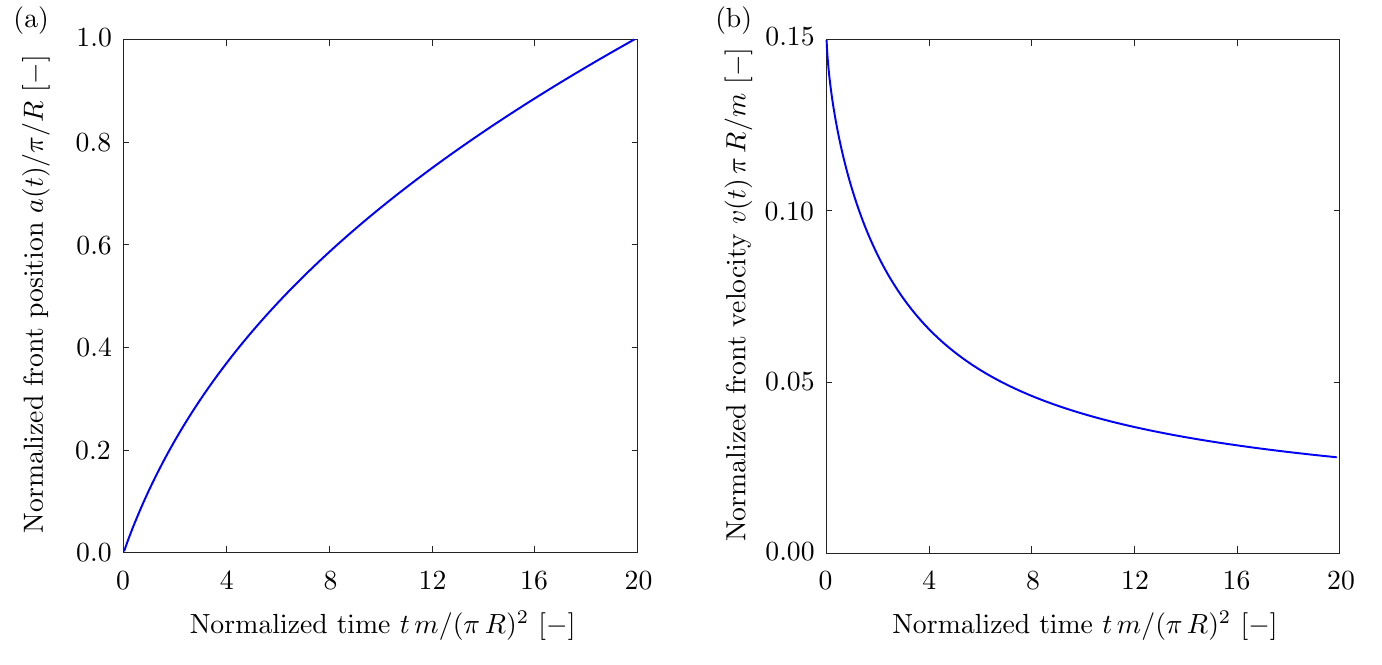}
	\caption{Normalized  front position (a) and  normalized front velocity (b) versus  scaled time.}
	\label{fig:P1_15}
\end{figure}

The position of the front and its velocity are two characteristic indicators  of the viral entry, giving insight into the current state of the process and enabling  the estimation of its total duration. The  evolution of these indicators (Fig. \ref{fig:P1_15}) shows that the front advances continuously during the process, whereas its velocity decreases with the strongest decline in the beginning, and an almost constant value at the end of the simulation. Similar behavior is shown in the work by Freund und Lin \cite{fre1}. However, while the overall trends in both works are the same, some interesting phenomena can be identified with regard to the limiting behavior. The current model experiences a weaker decline in the velocity resulting in an almost constant velocity towards the end of the simulation, which yields a linear advancing of the front. However, the velocity in \cite{fre1} moves towards zero which causes the process to experience almost no progress towards the end.

\subsection{Virus radius}
\begin{figure}
	\centering
	\includegraphics[width = \textwidth]{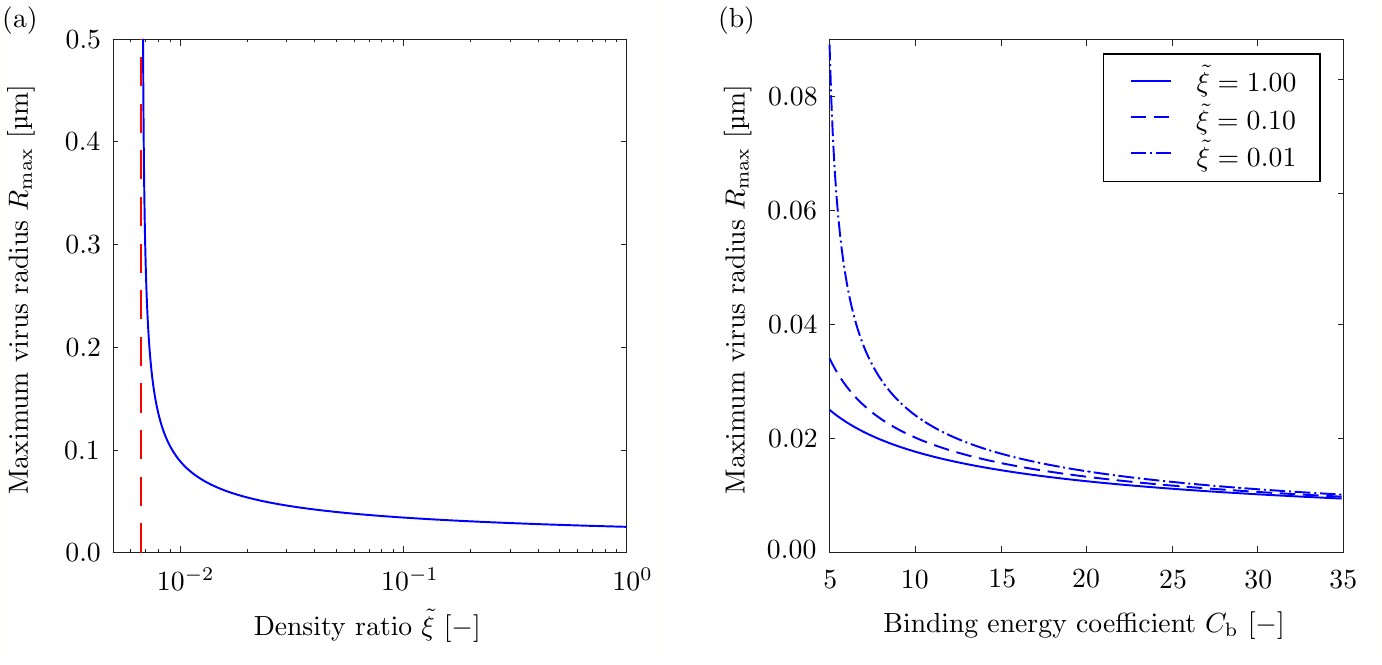}
	\caption{(a) Maximum virus radius  versus  receptor density ratio $\tilde{\xi}$. The vertical dashed line represents the critical value $\tilde{\xi}_\text{crit}$. (b) The maximum virus radius  versus the binding energy coefficient $C_\text{b}$ for different receptor density ratios $\tilde{\xi}$.}
	\label{fig:P1_r}
\end{figure}

Apart from the front position and velocity,  the virus radius also gives important information on the process, especially with regard to its initiation. This part of the analysis relies on the consequences of the energy balance \eqref{ic2}.
Since the right-hand side of this equation represents the kinetic energy, it  directly follows that the difference between the energy behind and ahead of the front on the left-hand side has to be non-negative. In the beginning of the process, when the density distribution is uniform and the front has not yet been established, the energy ahead of the front does not contribute to the total amount. Therefore, the part of the energy behind the front can be seen as an initial barrier that must be overcome in order to start the process. A study of the limiting case, where the front velocity approaches to zero, yields the expression  for the maximal radius
\begin{linenomath*}
	\begin{equation}
		\label{radius}
		R_\text{max} = \left.\sqrt{\frac{B}{2}}\right/\sqrt{\xi_{\text{eq}}\, C_\text{b} - \xi_{\text{eq}}\text{ln}\left(\frac{\xi_{\text{eq}}}{\xi_{0}}\right)}
	\end{equation}
\end{linenomath*}
under the condition that 
\begin{equation}\label{condition}
	C_\text{b} - \text{ln}\left(\frac{\xi_{\text{eq}}}{\xi_{0}}\right)> 0.
\end{equation}
By assuming the short notation for the receptor density ratio $\tilde{\xi} = \frac{\xi_{0}}{\xi_{\text{eq}}}$, the value for the critical density ratio results in
\begin{linenomath*}
	\begin{equation}
		\label{xitilde}
		\tilde{\xi}_\text{crit} = e^{-C_\text{b}}.
	\end{equation}
\end{linenomath*}
It is important to mention that the receptor density ratio is limited from both sides. On one hand, it holds  $\tilde{\xi} = \frac{\xi_{0}}{\xi_{\text{eq}}}\leq 1$ since $\xi_0\leq{\xi_{\text{eq}}}$. On the other  hand, it holds  $\tilde{\xi}>	\tilde{\xi}_\text{crit}$ due to the condition \eqref{condition}. The expressions \eqref{radius} and \eqref{xitilde} are now used  to study the values of maximum radius. These results are shown in Fig. \ref{fig:P1_r}a, where the  receptor density ratio is varied in the  admissible semi-open range $(\tilde{\xi}_\text{crit},1]$. Here, the 
red dashed line indicates the critical value $\tilde{\xi}_\text{crit}$ according to Eq. \eqref{xitilde}. The correlation between the radius and the density ratio has also been studied in the works by Gao et al.  \cite{gao2005}. They derive similar expressions for the limiting radius and the critical density ratio, however, their model provides  the expression for the minimal radius. In the present study as well as in \cite{gao2005}, the maximum radius  increases rapidly as $\tilde{\xi}$ approaches its critical value. This quantity \eqref{radius} also depends on the binding energy coefficient $C_\text{b}$ which is studied in  Fig. \ref{fig:P1_r}b  for three different values of the receptor density ratio. Here, the lower line corresponds to the higher ratio ($\tilde{\xi} = 1$), and the upper line to the lower ratio ($\tilde{\xi} = 0.01$). Whereas a noticeable difference of the radius is to be seen for smaller values of $C_\text{b}$, which becomes less pronounced as its value increases.

\subsection{Entry duration}
\begin{figure}
	\centering
	\includegraphics[width = \textwidth]{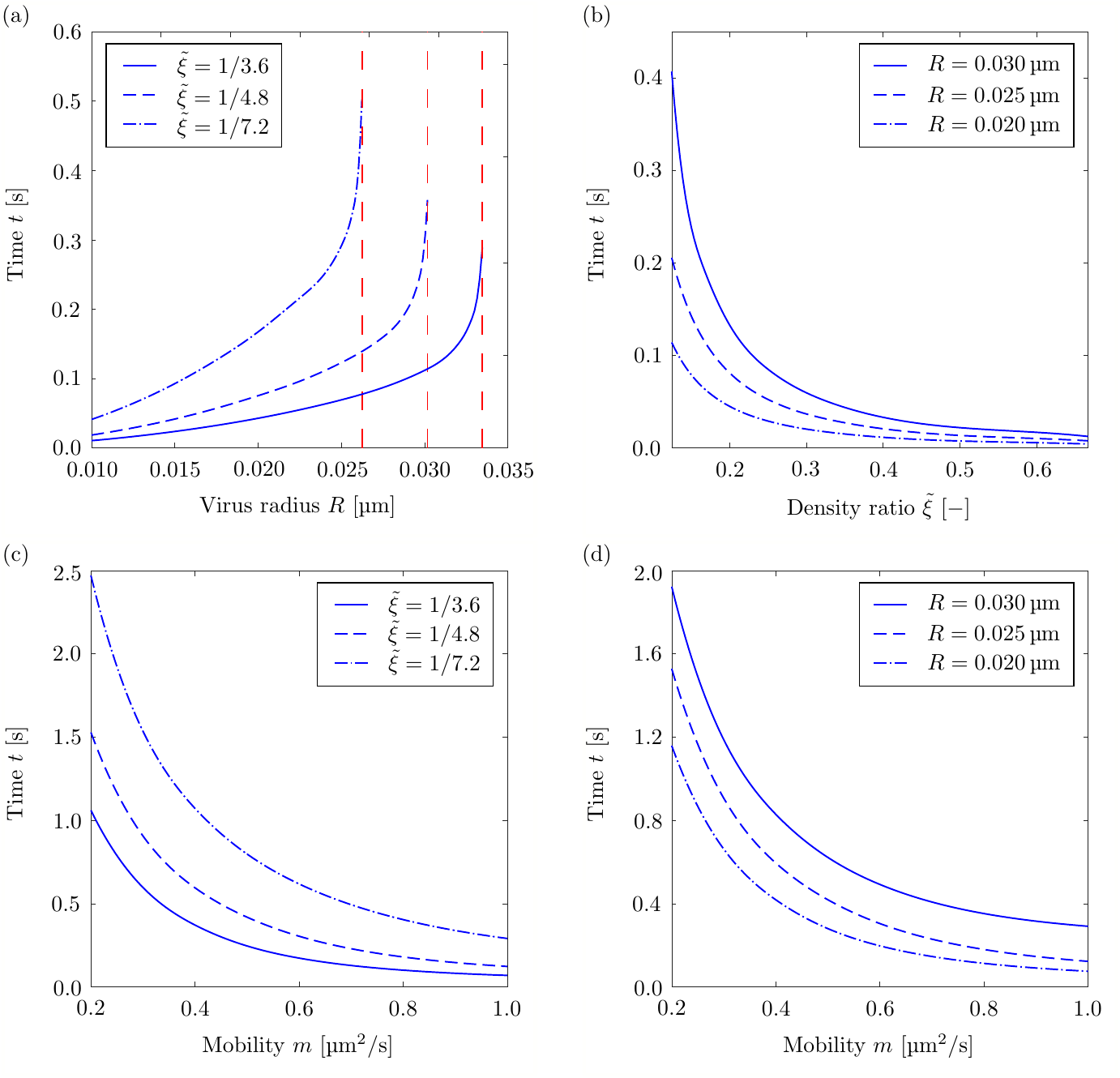}
	\caption{(a) Duration of the process depending on the virus radius for different  receptor density ratios $\tilde{\xi}$. Vertical dashed lines indicate the maximum radii for particular  critical receptor density ratios. (b) Duration of the process dependent on the receptor density ratio $\tilde{\xi}$ for different virus size $R$. (c) Process duration versus mobility parameter $m$  for  different receptor density ratios $\tilde{\xi}$. (d) Process duration  versus mobility $m$ for  different virus size $R$.}
	\label{fig:P1_t}
\end{figure}

A significant aspect of the  virus entry is the duration of the complete process as well as its dependence on different process parameters. The viral uptake via endocytosis ranges through different time scales. Ultra-fast-endocytosis takes 50-300 ms \cite{watanabe2013}, while kiss-and-run-endocytosis takes approximately 1 s \cite{zhang2009}. In the presented model, several parameters have a significant influence on the required time for the process. Three major parameters are the radius of the virus, the receptor density ratio and the mobility of the receptors. Figure \ref{fig:P1_t} shows the influence which each of these parameters has on the required time. The influence of the radius (Fig. \ref{fig:P1_t}a) is analyzed for three different density ratios. The red dashed lines correspond to the maximum radius determined according to Eq. \eqref{radius}. All curves show an increasing  trend with the highest value for the maximum radius. The curve with a smaller  value for $\tilde{\xi}$ indicates an increase in time, due to the higher difference in the receptor density available and required. 
In Fig. \ref{fig:P1_t}b, the time depending on the density ratio for  different radii is shown.  A longer process time is required for larger radii. The difference between the curves is small for larger ratios and becomes more significant as the ratio becomes smaller.

Finally, Fig. \ref{fig:P1_t}c and \ref{fig:P1_t}d both show the influence of the mobility on the required time.  The curves in Fig. \ref{fig:P1_t}c corresponds to different density ratios and indicate that lower density ratios are related to the higher time requirements. The difference between the curves for the different densities is more pronounced as the mobility takes smaller values. A similar  behavior is presented in Fig. \ref{fig:P1_t}d, where the  curves correspond to different radii. In both cases a decrease in the mobility causes an increase in the required time, which can be expected since a  larger $m$ enables the receptors to move more rapidly to the adhesion zone. An analysis of the influence of the radius and of the density ratio to the required time is also presented in the works by Gao et al. \cite{gao2005}. This research group also shows a strong increase in the required time for an increasing radius but proposes a minimum value for the radius. Furthermore, contribution \cite{gao2005} shows a rapid increase in the required time when the radius comes close to the minimum value, whereas the current model does not predict such a behavior. Similar to the results which are shown in Fig. \ref{fig:P1_t}b, Gao et al. \cite{gao2005} observe a strong decrease in the required time for decreasing density ratios with an upper limit at $\tilde{\xi} = 1$.

\subsection{Cooperativity}
\begin{figure}
	\centering
	\includegraphics[width = \textwidth]{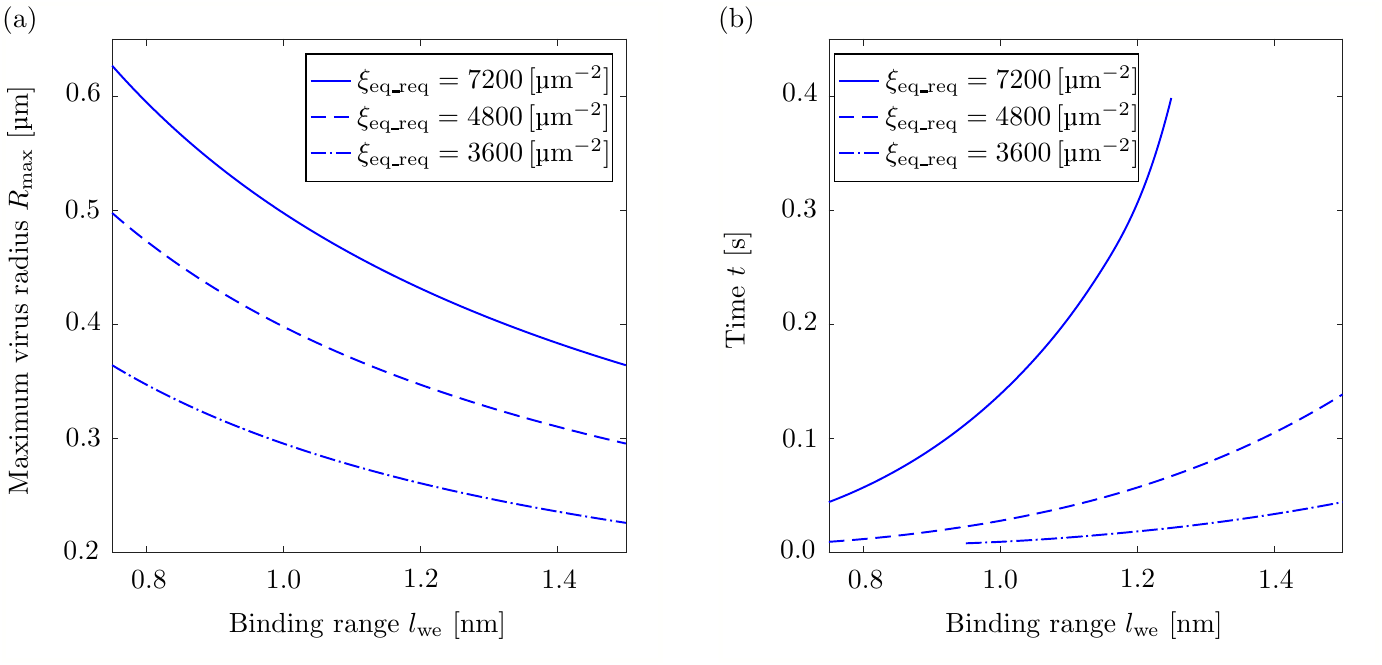}
	\caption{(a) Maximal virus radius depending on the binding range $l_\text{we}$ for given required receptor density $\xi_{\text{eq\_req}}$.  (b) Process duration depending on the binding range $l_\text{we}$ for given required receptor density $\xi_{\text{eq\_req}}$.}
	\label{fig:lw}
\end{figure}

A more  comprehensive study  of the uptake process also requires the  data on cooperativity (Sec. \ref{Cooperativity}) to be included in the model. The influence of this factor is demonstrated on the basis of two examples dealing with the  effects of the binding range $l_{\text{we}}$.
The required receptor density Eq. \eqref{xireq} in combination with the equation for the radius Eq. \eqref{radius} provides a relation between the binding range and the limiting radius. The corresponding results  are presented in Fig. \ref{fig:lw}a where the single curves are related to different receptor densities of the virus. As already  shown in Fig.~\ref{fig:P1_7}b, a smaller receptor density of the virus benefits the process, resulting in a larger possible radius.

Furthermore,  the influence of the binding range $l_{\text{we}}$  on the required density \eqref{xireq} and indirectly on the duration of the process is presented in Fig. \ref{fig:lw}b. Again, the three curves correspond to different receptor densities of the virus. For the upper curve, corresponding to $\xi_{\text{eq}} = 7200\,\,\si{\micro}$m$^{-2}$, a  large binding range hinders the begin of the process, since  the required density $\xi_{\text{eq\_req}}$ becomes to be too high. On the contrary, the lower curve corresponding to $\xi_{\text{eq}} = 3600\,\,\si{\micro}$m$^{-2}$, shows a lower threshold for the process to take place. Here, the duration of the process strongly depends on the binding range itself and prefers lower values in order to complete the process quickly. However, for an extremely small $l_{\text{we}}$, the number of virus and cell receptors coming into contact is not sufficient and the process cannot start. 

\subsection{Interaction of selected parameters}
\begin{figure}
	\centering
	\includegraphics[width = \textwidth]{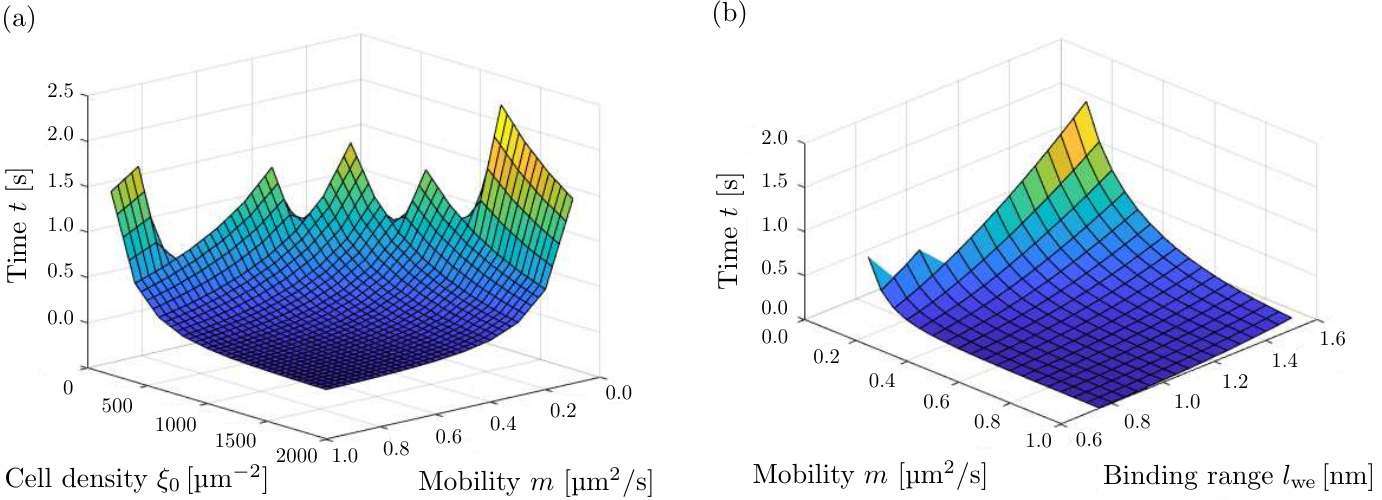}
	\caption{(a) Process duration depending on  mobility $m$ and the initial receptor density $\xi_0$. (b) Process duration depending on  mobility $m$ and the binding range $l_\text{we}$. Plots also show the admissibility ranges for chosen parameter sets.}
	\label{fig:mt}
\end{figure}

The discussion of results closes by presenting the interaction of selected process parameters  and their influence on the initiation   and duration of the process. Figure \ref{fig:mt}a shows the  combination of  the initial receptor density $\xi_{0}$ and  mobility $m$, and shows a strong increase in the required time for the parameters chosen. If both parameters take small values, the process does not start, but once this threshold is surpassed the required time drops rapidly regardless of which parameter is changed. Especially the area with small values for the mobility is interesting, since it is not uncommon for viruses to attach to cell receptors with small mobilities.  Some typical examples are the HIV-virus connecting to a receptor  with mobility  $m = 0.05\,\,\si{\micro}$m$^2/$s \cite{garcia2020} or the Semliki Forest virus connecting to a receptor with $m = 0.01\,\,\si{\micro}$m$^2/$s \cite{gao2005}. 
Finally, Fig. \ref{fig:mt}b shows a range of combinations for  mobility $m$ and binding range $l_{\text{we}}$ in which the process takes place. The influence of the binding range on the required time is weak, compared to the influence of the mobility, and becomes more noticeable for smaller values of the mobility.

\section{Conclusion and outlook}
\label{Conclusion and outlook}
The present study focuses on the investigation of the viral entry driven by the receptor diffusion using the finite difference method as simulation technique.
An approach based on the consideration of the energetic aspects yields a formulation providing a well-posed description of the endocytosis process. The motion of the receptors is described by the diffusion differential equation accompanied by two boundary conditions dealing  with the flux balance at the ends of the considered area. In addition, a supplementary condition is introduced to define the energy balance at the adhesion front.

The model  developed  shows  several important features: the definition of the supplementary  condition only depends on the quantities at the front, and the numerical simulation  of the problem bypasses the introduction  of assumptions typical of  an analytical solution. The approach is highly efficient with regard to time and computer capacity, such that a fast simulation of different scenarios and a profound study of process parameters are possible. Here, the influences  and the interaction of mobility, receptor densities, virus size and receptor cooperativity play a central role. Their analysis, for example, yields data  on the admissible regions, the upper limit of the size of the virus  able to enter the cell and the estimation of the process duration. Amongst others, the analysis shows that the process duration strongly increases when   a virus size approaches a critical value and that extremly high and low values of binding range have an impeding influence on the process initiation.

The results presented in this work pertain to a helical and a spherical virus penetrating a flat cell surface, which enables the taking of advantage of the axial and rotational symmetry and perform simulations in a two dimensional setup. However, an extension to a three dimensional setup has to be taken into account in order to analyze the receptor distribution for a non-spherical virus or a non-homogeneous receptor density of the cell. Furthermore, additional contributions, for example, caused by bending of the cell ahead of the front, can be considered in the energetic  supplementary condition. Alternative expressions for bending lipid bilayers can also be introduced in order to carry out  more realistic simulations.

\section*{Acknowledgments}
We gratefully acknowledge the financial support of the German Research Foundation (DFG), research grant No. KL
2678/7-1, and the Austrian Science Fund (FWF), research grant No. I 3431-N32.
We also thank Dr. Matias Hernandez from the Max Planck Institute of Molecular Physiology at Dortmund, Germany, for valuable discussions.

\bibliography{references}

\end{document}